# Advancements in Real-Time Oncology Diagnosis: Harnessing AI and Image Fusion Techniques

Leila Bagheriye, and Johan Kwisthout

*Abstract*—**Real-time computer-aided diagnosis using artificial intelligence (AI), with images, can help oncologists diagnose cancer with high accuracy and in an early phase. We reviewed real-time AI-based analyzed images for decision-making in different cancer types. This paper provides insights into the present and future potential of real-time imaging and image fusion. It explores various real-time techniques, encompassing technical solutions, AI-based imaging, and image fusion diagnosis across multiple anatomical areas, and electromagnetic needle tracking. To provide a thorough overview, this paper discusses ultrasound image fusion, real-time in vivo cancer diagnosis with different spectroscopic techniques, different real-time optical imaging-based cancer diagnosis techniques, elastography-based cancer diagnosis, cervical cancer detection using neuromorphic architectures, different fluorescence image-based cancer diagnosis techniques, and hyperspectral imaging-based cancer diagnosis. We close by offering a more futuristic overview to solve existing problems in real-time image-based cancer diagnosis**.

*Index Terms*—**real-time, image fusion, cancer, diagnosis, Artificial intelligence, spectroscopy, hardware accelerator, Hyperspectral imaging.**

## I. INTRODUCTION

REAL-TIME medical image-based cancer diagnosis utilizing artificial intelligence (AI) has recently gained widespread use in oncology. It has been shown that real-time imaging improves the diagnosis, treatment, and follow-up process of patients. Studies have demonstrated that in order to achieve a satisfactory patient-specific diagnostic and therapeutic process combining pretreatment clinical variables with computer-aided intraoperative imaging information is essential [1]. Image fusion shows considerable potential in aiding tumor diagnosis, directing biopsies, and facilitating interventional ablations, adding accuracy in the operation room [2].

In [1], intraoperative real-time navigation systems using transrectal ultrasonography (TRUS) and magnetic resonance imaging (MRI) are employed to achieve a more precise spatial cancer mapping and interventional targeting therapy. In [2], the fusion of endoscopic ultrasound computed tomography (EUS-CT) images aided in navigating and characterizing the target tumor and its adjacent anatomical features. The process of image fusion, which involves directly comparing the target lesion with other images, helps multidisciplinary teams achieve clearer visualization for surgical or radiation treatment planning. MRI-ultrasound (US) image fusion has been utilized for prostate biopsy using the ARTEMIS semi-robotic fusion biopsy system [3]. MR-US technology, as a multiparametric approach along with semi-robotic, electronic-mechanical tracking of the prostate, helps clinicians precisely navigate needles to identify lesions, subsequently improving the accuracy of the biopsy. The system utilizes motion compensation to align previously constructed virtual images with real-time US images as the physician navigates the needle to a target using the probe and mechanical arm. This method enhances the ability to target regions of interest (ROIs) and, in systematic biopsies, helps prevent both under-sampling and over-sampling. The system records the precise locations of the taken cores. Furthermore, it can import previously recorded patient information, providing valuable data for future biopsies. Endogenous molecular-specific computed tomography (CT) and MRI offer three-dimensional anatomical images, proving effective in identifying late-stage or large tumors. Sophisticated hybrid imaging methods like positron emission tomography (PET)/CT can conduct molecular imaging by utilizing exogenous biomarkers [4]. Magnetic resonance spectroscopy offers detailed insights into tissue biochemistry, yet its clinical utility is restricted. For examining epithelial tissues (where more than 80% of primary cancers arise), these traditional medical imaging methods are not optimal [5], [6].

Optical imaging and spectroscopic techniques, while having limited penetration, are well-suited for examining surface epithelial tissues for early cancer detection. There are optical techniques such as diffuse reflectance spectroscopy (DRS), autofluorescence spectroscopy (AFS), and autofluorescence imaging which provide non-destructive and exogenous agent-free tissue sensing [7]. While the sensitivity of intraepithelial neoplasia detection can be improved by autofluorescence imaging, DRS, and AFS can significantly improve, their diagnostic specificities are moderate [8]. For the decreased specificity, it can be partly explained with the spectral similarities between precancerous tissues and benign abnormalities, such as inflammation, can partly account. This

 Leila Bagheriye and Johan Kwisthout are embedded in the Donders Institute for Brain, Cognition and Behaviour, Radboud University, Nijmegen, P.O. Box 9104, 6500HE Nijmegen, Netherlands. (e-mail: Leila.bagheriye@donders.ru.nl; Johan.Kwisthout@donders.ru.nl ).

similarity could also pose a significant challenge for the successful clinical adoption of DRS and AFS. [9], [10]. On the other hand, various AI techniques have been employed for cancer diagnosis within real-time imaging/image fusion, but the lack of external validation poses a significant obstacle to the secure and regular utilization of AI classification models in clinical practice [11]. Therefore, to take an essential step toward the clinical application of this technology, the classification model needs external validation on a separate cohort of patients.

This paper provides an overview of the current literature regarding AI-based solutions for imaging in real-time and image fusion. Section II discusses US-based imaging and image fusion. Section III introduces real-time in vivo cancer diagnosis with different spectroscopy techniques. Section IV describes different real-time optical imaging-based cancer diagnosis methods. Section V explains elastography-based cancer diagnosis methods. Section VII describes different fluorescence image-based cancer diagnosis techniques. Hyperspectral imaging-based cancer diagnosis is covered in section VIII. Section IX describes PET/MRI image deep learning for breast cancer. In section X, we provide an overall discussion of the different approaches and future work. Finally, section XI concludes the paper.

## II. Ultrasound Image Fusion

In this section, US-based imaging will be covered. Online US-MRI-based image fusion, real-time transabdominal and endoscopic US, imaging of liver US, as well as speckle reduction of US images, will be described.

### A. MRI-US Image Fusion of Prostate Biopsy

To enhance the precision of prostate biopsy, various approaches can be considered. These include merging US images with MRI, integrating contrast-enhanced US, utilizing elastography, combining 3D US with computer-assisted techniques and utilizing Robotic baes needle guidance [1]. TRUS imaging has primarily served as a basic guidance tool for systematic biopsy. Its role is to ensure that biopsies are taken from all sextant areas of the prostate in a random manner. Additionally, a small number of biopsies are taken by targeting suspicious lesions directly. The fusion of real-time TRUS with preoperative MRI has been introduced (Fig. 1). The goal is to compensate for the limitations of current 2D grayscale TRUS alone for intraoperative guidance to enhance the precision of targeted prostate biopsy and prostate intervention [1]. In Fig. 1(a), no detectable lesion is shown in the real-time axial TRUS image, while Fig. 1(b) on the right side shows the lesion with the synchronized, previously acquired MR image. In Fig. 1(c), the MR is overlayed onto the TRUS (image fusion), aiding in the targeting of the lesion visible only on the MRI. Fig. 1(d) shows real-time TRUS visualization of the hyperechoic needle, using a needle tracking system similar to a global positioning system (GPS), allowing precise documentation of the exact location of each biopsy sample with spatial digital data. In Fig. 1(e), digital marks were used to register the spatial points of both the distal and proximal endpoints of the biopsy trajectory in the 3D MR volume data of the phantom. Fig. 1(f) shows fusion with GPS-like technology. In Fig. 1(g), Later on, real-time TRUS guidance is utilized for spatial re-targeting. Fig. 1(f) illustrates the ability to revisit a previously marked biopsy location using newly acquired 3D volume data. This is achieved through computer-based marking, similar to GPS, of the previous biopsy position.

Moreover, Fig. 1(h-k) shows augmented reality navigation in laparoscopic radical prostatectomy. Augmented reality integrates intraoperative surgical navigation to overlay a 3D image onto the live surgical view. This technology displays 3D anatomy beyond the surgical view, revealing the anatomical orientation of the targeted pathology and surrounding tissues before surgical exposure []. Augmented reality involves technical steps including image acquisition, segmentation, registration, visualization, and navigation. These can be integrated with energy-based ablative machines and robotics. Fig. 2(h) and (i) show views from different angles. The workstation created a 3D surgical model of the prostate from original intraoperatively acquired real-time 2D transrectal ultrasonography imagery, demonstrating the area of biopsy-proven cancer, indicated by a blue color (which is located on the left-posterior-lateral surface of the prostate). As shown in Fig. 2(j), via augmented reality techniques, the 3D model of the prostate and seminal vesicles are overlaid onto the real endoscopic view. Fig. 2(k) shows the 3D model of the cancerous lesion overlaid on the left-posterior-lateral aspect of the prostate. In [12], for augmented reality overlay a stereo-endoscopic visualization system has been described. Without the need for external tracking devices it allows 3D-to-3D registration of the preoperative CT scan with the real surgical field. As a result, the integration of stereoscopic vision into robotic systems represents a significant advancement in augmenting surgeons human senses during image-guided surgery.

### B. Real-Time Online Image Fusion with Transabdominal/ Endoscopic US

To enhance the diagnosis and interventional treatment of focal liver lesions, real-time US fusion for the liver is emerging as a promising approach over traditional US methods. Image fusion provides considerable benefits for pinpointing liver lesions in minimally invasive procedures like biopsies, percutaneous ablations, or planning radiation treatments [2]. Simpler navigation and characterization of the target tumor and nearby anatomical structures can be achieved by combining EUS and CT images.

A typical image fusion session, as shown in Fig. 2(a), which is composed of the electromagnetic (EM) field generator, external physical markers, and the image fusion software [2]. The EM tracking and field generator system is positioned near the patient and connected to the computer intended for running the fusion imaging software. Then active marker disks are then placed on the patient's xiphoid process. The EM sensor is inserted into the navigation catheter within the working channel of the endoscope or echoendoscope [2]. In order to generate a 3D model of the patient's anatomy, pre-procedure CT scans are imported into the fusion imaging (FI) software. For identifying the target, dual visualization of the EUS image and its corresponding virtual section through the CT volume are employed.

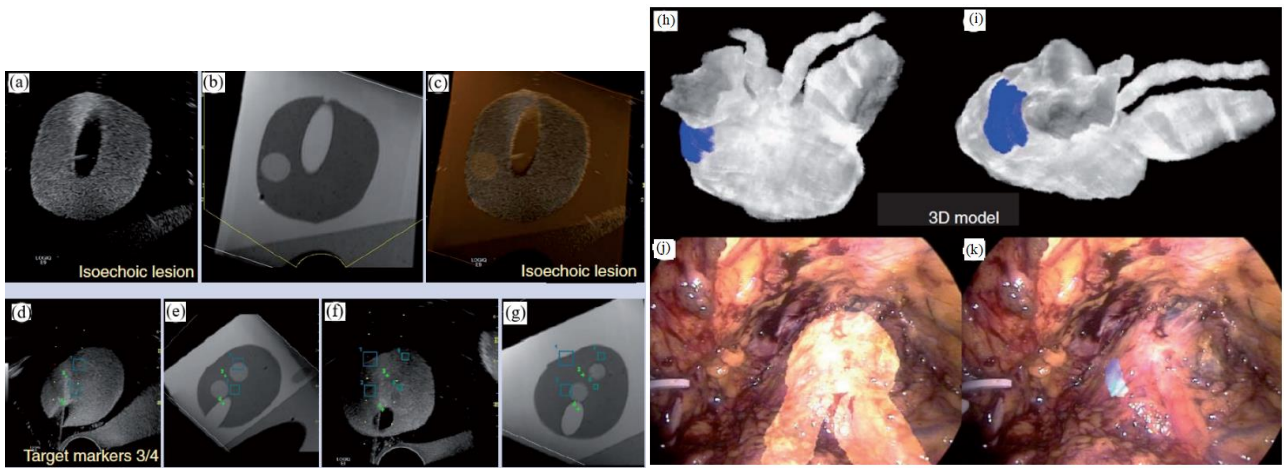

Fig. 1. Fusion platform for hybridizing transrectal ultrasonography (TRUS) with magnetic resonance imaging (MRI). (a) For axial TRUS image in real-time mode there is no detectable lesion. (b) The lesion on the right side is visualized with the synchronized, previously acquired MR image. (c) The MR overlays onto the TRUS as image fusion to facilitate the determination of the lesion which is visible only on the MR. (d) Real-time TRUS visualization of the hyperechoic needle (with a GPS-like needle tracking system) documents with spatial digital data the precise location of each biopsy sample. (e) Digital marks are used to register the spatial coordinates of both the distal and proximal endpoints of the biopsy trajectory. (f) Subsequently, for spatial re-targeting, real-time TRUS guidance is employed. (g) With GPS-like computer-based marking of the previous biopsy position biopsy location is determined using newly acquired 3D volume data [1]. Augmented reality navigation in laparoscopic radical prostatectomy: (h) and (i) show views from different angles. In order to determine the blue-colored cancer area (which is proved with biopsy) on the left-posterior-lateral surface of the prostate, intraoperatively, the workstation reconstructed a 3D prostate surgical model. (j) The 3D model of the prostate and seminal vesicles are overlayed onto the real endoscopic view with augmented reality techniques. (k) the cancer lesion is overlayed on the left-posterior-lateral prostate (the 3D model) [1].

After reaching the target, the EUS position is fixed. To collect a biopsy, the navigation catheter is withdrawn and substituted with a fine aspiration needle [2].

Fig. 2(b-k) show the software platform for EUS/CT image fusion, which provide automatic processing of CT for 3D rendering and segmentation, nodule detection, organ/tumor segmentation, registration of patient's CT for endoscopy procedures, localization and tracing of therapeutic devices using tracking technologies, virtual visualization of medical instruments on the CT stack, and augmented reality for virtual visualization of the patient's anatomy over intraprocedural video [2]. An early version of the software underwent testing using a specially constructed model of pig organs.

At first, a CT scan of the model was conducted using a specific pancreatic protocol. The images were then segmented, and 3D reconstruction was done before uploading into the software. This enabled the visualization of the echoendoscope's position within the 3D CT cube and real-time movement of the same section.

The integration of real-time EUS with CT significantly improved the visualization of specific lesions or organs, enabling complex therapeutic procedures and improving operator confidence. Fig. 2(b) (shows image fusion (EUS/CT) testing), shows the 3D reconstruction of the segmented CT scan, featuring a simulated gallbladder with stones, a pseudocyst, and the liver, illustrating real-time EUS-CT fusion. The oblique section displays the co-registered large-field CT, showing all three organs, alongside the narrow EUS image, which only shows the pseudocyst.

*C. Real-Time fusion imaging of liver US*

US imaging of the liver is frequently employed in clinical practice for diagnosing and treating liver disease, providing various liver images [13]. However, inconsistencies exist between CT scans, MRIs, and liver US images when evaluating focal liver lesions, as CT or MRI scans provide cross-sectional views, including coronal and sagittal plane images. For accurate identification of the areas of interest in these images, it is necessary to mentally align CT or MRI images with the planes corresponding to real-time B-mode ultrasound locations. While liver US does not require additional instruments, considerable training and experience are needed, as operator errors can occur.

Air in the lungs and intestinal gas present during liver ultrasound scans may obstruct the acoustic window, possibly masking small or inconspicuous foci in difficult-to-observe regions. The initial steps in real-time US fusion imaging of the liver entail preparing and positioning the patient, along with deploying electromagnetic field equipment and a US machine. In order to achieve high accuracy in comparisons during real-time US fusion imaging, it is necessary for the patient to maintain a relaxed posture to minimize movements. Image fusion begins with retrieving stored CT or MRI images. By initiating the electromagnetic field generator, the US machine acquires dynamic US images. The spatial information in the CT or MRI images and US images is adjusted based on the positioning point, plane, and 3D data to merge the two sets of images and display them in real-time. The fusion US is then carried out.

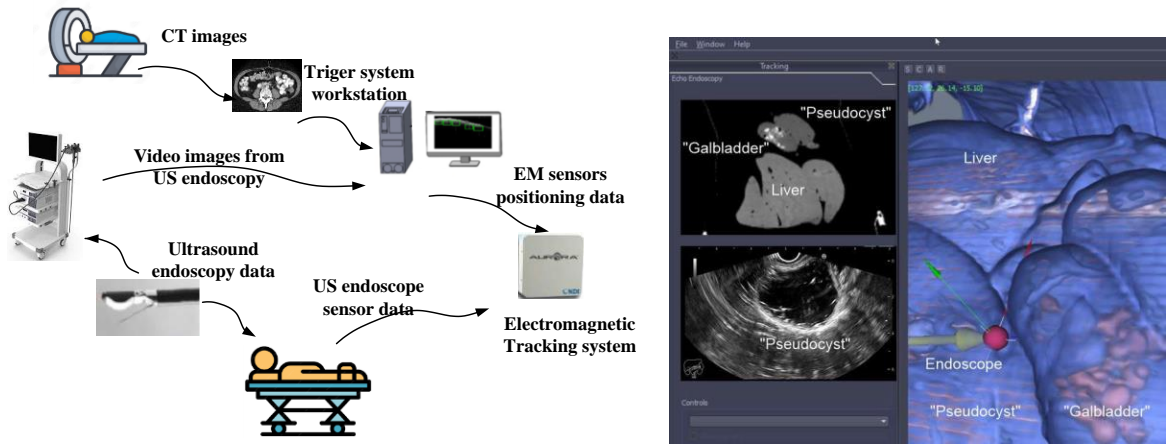

Fig. 2 (a). A typical image fusion framework. A virtual 3D map of the patient is created based on pre-procedure CT images. Subsequently in order to overlay the EM sensor locations onto the 3D model, the image fusion software is used. To identify the clinical target, live EUS images are merged with the 3D model [2]. Fig. 2(b) shows the real-time image fusion (EUS/CT) through which the 3D reconstruction of the segmented CT has been done (featuring a phantom 'gallbladder with stones,' 'pseudocyst,' and liver). The oblique section represents the co-registered large-field CT, which is showing all three organs. The narrow EUS image, shows the pseudocyst (adapted from [2]).

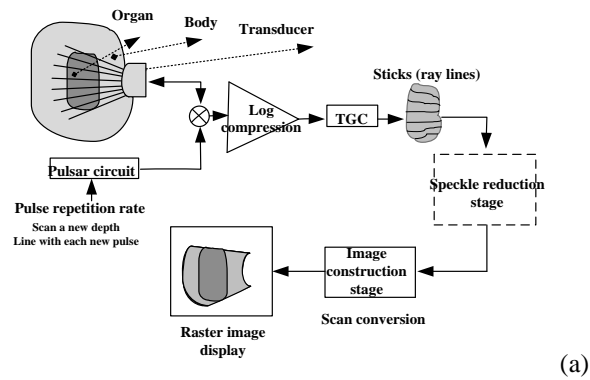

(a)

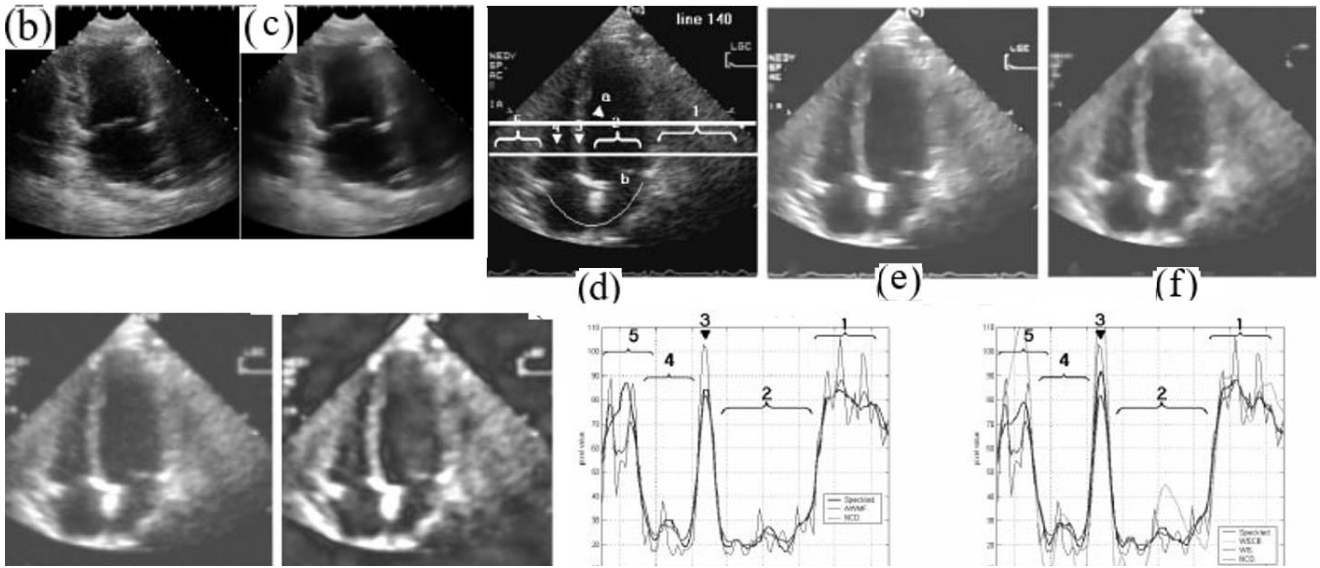

Fig. 3 (a) The placement of the speckle reduction stage which is shown via the block diagram of the convex B-mode ultrasonic system [14]. (b). Heart short-axis view. (c) The original image. (d) Displayed on the right is the image after noise reduction using nonlinear coherent diffusion model (NCD). (e) Original speckled image with a superimposed line mark. (f)-(i) Images processed with NCD, adaptive weighted median filter (AWMF), wavelet shrinkage (WS), and model and the wavelet shrinkage coherence enhancing (WSCE) methods, respectively [14]. (j) and (k) Evolution profiles of the row line marked in (c) using different algorithms. [14].

The operator positions the US probe to capture images in a plane that corresponds to one of the previously acquired CT or MRI images, usually in a cross-sectional plane, and designates this plane as the matching coordinates. The matching process relies on 3D information, where a computer compares and aligns the captured data with the 3D-reconstructed spatial coordinates derived from CT or MRI images.

### D. US speckle reduction method

To reduce speckle noise in US images while preserving structured regions and organ surfaces, a nonlinear coherent diffusion model (NCD) was proposed [14]. This technique offers advantages such as reliable parameter selection, fast computation, and maintenance of texture and organ surfaces. The NCD model shows potential to improve real-time US imaging and aid automated segmentation techniques.

The NCD model adapts to speckle extent and image anisotropy, transitioning from isotropic diffusion to anisotropic coherent diffusion and finally to mean curvature motion [14]. This progression effectively preserves information related to resolved-object structures while filtering out fully developed speckles in the image. A block diagram of the ultrasonic convex B-mode system is shown in Fig. 3(a). In this figure, the proposed speckle reduction stage is applied to scan lines instead of the constructed raster image.

In the suggested algorithm, a streamlined discretization method allows for immediate application on standard commercial systems. The technique involves smoothing log-compressed US pulse-echo images to suppress fully formed speckles (FFS). To this end, a nonlinear diffusion model, is adapted to remove the compressed speckle pattern from raw scan lines.

Using raw scan lines rather than the composed image yields greater accuracy. This is because real scans are represented in radial coordinates, not cartesian coordinates. This approach is nearly ten times faster, compared to its application on the formed image. The technique's anisotropic nature enables the enhancement of coherent structures, with the diffusion model's dynamics controlled by the local signal behavior. To regulate both the direction and intensity of diffusion, the model evaluates signal coherence, a crucial characteristic of speckle.

In order to achieve real-time implementation, a unique discretization scheme is suggested to minimize method intricacy. Computer simulations and real US images are used to analyze and validate the stability and robustness of these techniques.

Several approaches have been suggested to tackle the issue of speckle noise. Wavelet-based methods involve converting the image into cartesian space, leading to increased computational time and diminished solution accuracy. Furthermore, the adaptive weighted median filter (AWMF) model fails to fully account for the 2-D nature of the image.

On the other hand, the mean curvature motion (MCM) method was developed to improve piecewise constant images, a feature not typically prevalent in ultrasound imaging. Furthermore, this technique is prone to noise sensitivity because it employs second-order derivatives to estimate the mean curvature direction, potentially leading to significant degradation in results.

The suggested model imitates MCM solely in proximity to borders and areas with high specularity. The findings indicate that the method holds significant promise for supporting segmentation techniques and automated calculations of area/volume, such as measuring the area of heart chambers.

Despite effectively smoothing both cavities and muscular tissue, wavelet shrinkage (WS) and AWMF were unable to eliminate the speckle pattern within the heart cavity, resulting in blurred boundaries. The profile line indicates that the cavities (marks 2 and 4) are smoothed to the greatest extent, whereas structures (marked as 5) are most effectively preserved by NCD.

## III. REAL-TIME CANCER DIAGNOSIS USING VARIOUS SPECTROSCOPY TECHNIQUES

In this section, different spectroscopy technologies will be explained. To this end, real-time cancer diagnosis using Raman spectroscopy, cancer diagnosis using a handheld mass spectrometry system, gastrointestinal cancers using diffuse reflectance spectroscopy, and real-time endoscopic Raman spectroscopy will be described.

### A. Real-Time Cancer Diagnosis Through In Vivo Raman Spectroscopy

Histopathological analysis of tissue samples obtained through biopsy stands as the gold standard for diagnosing cancer. [4]. The diagnostic procedure is lengthy, causes anxiety, and extends the overall duration of the process. It's crucial to steer clear of unnecessary biopsies. However, patients cannot afford to overlook cancerous lesions that warrant biopsy. Accurately diagnosing cancer in real-time and in vivo, with high sensitivity and specificity while sparing most benign lesions, poses a significant challenge. For this purpose, Raman spectroscopy has emerged as a practical tool for rapid in vivo tissue diagnosis.

Subtle biochemical compositional changes are captured by Raman technique in the early stages of tissue malignancy, before any visible morphological changes occur. Without taking a biopsy, by directing a laser beam directly onto suspected cancerous lesions, in vivo Raman spectra are measured. Important developments such as proprietary Raman spectrometer designs enhance spectral signal-to-noise ratios and decrease the necessary integration time to mere seconds; The advancement of data mining and bioinformatics research significantly streamlines the automated analysis of Raman spectra.

Raman spectroscopy through providing high molecular specificity, effectively reduces the occurrence of false positive biopsies in cancer diagnosis. Fig. 4(a) 4(a) depicts a schematic of a real-time Raman spectrometer designed for skin cancer detection. Without correction, the image captured by the charge-coupled device (CCD) camera with a straight entrance slit exhibits curvature. Arranging the optic fiber array in a reverse orientation effectively corrects the curvature.

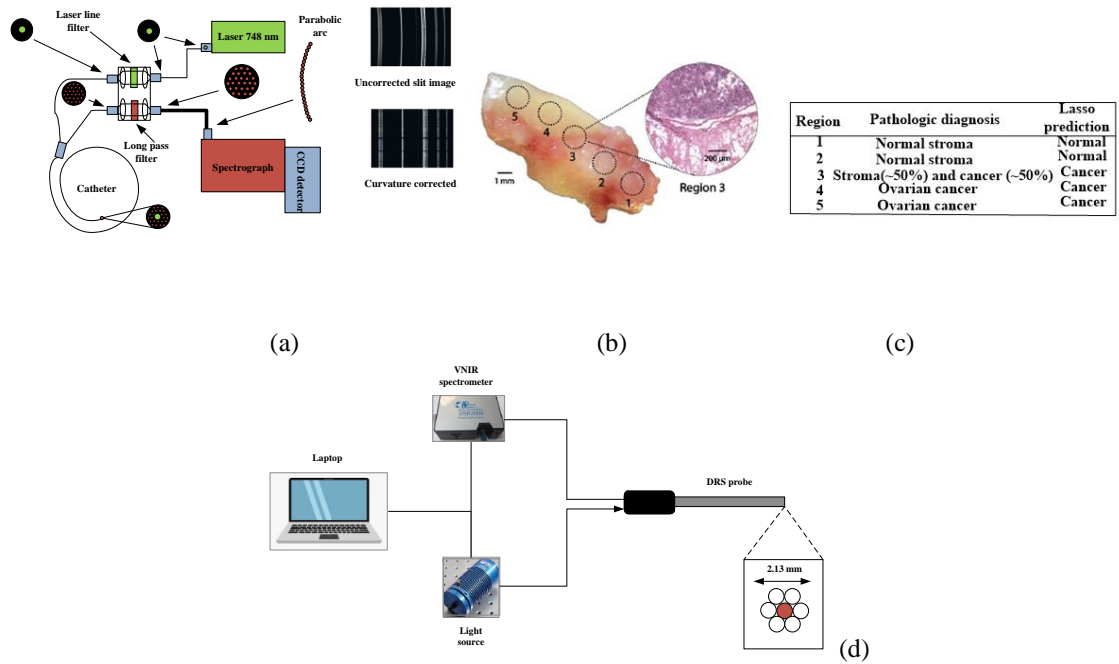

Fig. 4(a) Diagram illustrating the high-speed, real-time Raman system designed for diagnosing skin cancer ([4]). (b) A mixed histologic composition in an high-grade serous ovarian cancer (HGSC) tissue sample analyzed with the MasSpec Pen. (a) MasSpec Pen based analysis on optical image of the tissue sample at regions 1 to 5, followed by freezing, sectioning, and hematoxylin and eosin (H&E) staining. (b) The inset shows the H&E-stained tissue section from region 3. It reveals a mixed histologic composition that includes both cancer and adjacent normal stroma tissue. (c) the diagnosis of the five regions has been shown in Table alongside the Lasso prediction results ( [27]). (d) The setup and instrumentation for diffuse reflectance spectroscopy (DRS) for data acquisition and sample illumination. The DRS probe was connected to both the light source and the spectrometer. All electronic devices communicated with the software [1].

Different methods have been used in Raman analysis such as naïve Bayesian classifiers (NBC), linear discriminant analysis (LDA), support vector machine (SVM), and artificial neural network (ANN) [15]. Training a Bayesian classifier is straightforward, and it performs well even when the assumption of feature independence is not met [16].

Besides assuming conditional probability density functions, decision hyperplanes can be computed to partition the feature space into regions for each class. with Raman spectroscopy, LDA is a most popular technique for classifying malignant tissues [17, 18]. Generalized discriminant analysis (GDA) maps the original variables into a new feature space where the variables are nonlinearly correlated with the originals [19]. Both LDA and GDA are frequently employed together with principle component analysis (PCA) or partial least squares (PLS) [20, 21]. For the classification of cancerous tissues using Raman spectroscopy SVMs has been reported [22, 23]. For SVMs, optimal generalization performance is attained with high-dimensional data and/or datasets with a low ratio of training samples to input dimensionality [24]. When data follows a normal distribution, LDA typically offers good generalization compared to SVMs, which require tuning numerous parameters and may lead to overfitting [24]. Training neural network classifiers demands intensive computation and encounters issues such as converging to local minima and susceptibility to outliers. [25]. For the classification of Raman spectra for tissue characterization, neural networks are used rarely [26].

### B. handheld mass spectrometry system for cancer diagnosis

Traditional approaches to histopathological tissue diagnosis require significant labor and time, often causing delays in decision-making for diagnostic and therapeutic interventions. A recently reported advancement involves the creation of MasSpec Pen, a handheld mass spectrometry device designed for swift and noninvasive diagnosis of human cancer tissues, utilizing automation and biocompatible materials [27]. For efficient extraction of biomolecules, the MasSpec Pen enables precise and automated dispensing of a defined water droplet onto tissue surfaces. The MasSpec Pen was employed for ex vivo molecular analysis. It examines 20 thin tissue sections from human cancer specimens and 253 tissue samples from patients [27]. These samples comprised both normal and cancerous tissues from a range of organs, including the breast, lung, thyroid, and ovary.

The mass spectra acquired displayed rich molecular profiles containing a range of potential cancer biomarkers, such as metabolites, lipids, and proteins [27]. Statistical classifiers, based on the molecular data validated through histological methods, enabled accurate cancer predictions with high sensitivity, specificity, and overall accuracy. These classifiers were also capable of distinguishing between benign and malignant thyroid tumors, as well as identifying different histologic subtypes of lung cancer. The classifier enabled cancer diagnosis with high accuracy even in marginal tumor regions displaying mixed histologic composition.

It has been shown that the MasSpec Pen is applicable for in vivo cancer diagnosis during surgery in tumor-bearing mouse models,

accomplishing this without inducing any discernible tissue damage or stress to the animal. MasSpec Pen has been shown it's high potential to be employed as a clinical and intraoperative technology in ex vivo and in vivo cancer diagnosis [27].

To build molecular databases, averages of three mass spectra were used. The full mass range of the spectra was partitioned, and four representative mass spectra were analyzed for PCA for each tissue section. For tissue classification, the least absolute shrinkage and selection (LASSO) method was applied. Since LASSO yields "sparse" models, the generated models are are easier to interpret compared to other regularization methods. The LASSO method computes a mathematical weight for each statistically significant feature, determined by the feature's importance in characterizing a specific class in mass spectral analysis.

The dataset collected from the 253 human tissue samples underwent PCA analysis, revealing a clear distinction between cancerous and healthy tissues across various organs. Since the first three PCs explain 77%, 69%, 51%, and 87% of the total variance of breast, thyroid, lung, and ovarian data sets, respectively, PC1, PC2, and PC3 are used for analysis [27].

To pinpoint histologically unique areas within a solitary human tissue sample featuring high-grade serous ovarian cancer (HGSC) alongside normal ovarian stroma tissue, we assessed the efficacy of the MasSpec Pen. The MasSpec Pen sequentially analyzed five regions of the tissue sample, as determined in the optical image shown in Fig. 4(b). The regions were labeled from 1 to 5.

Following the MasSpec Pen analysis, the tissue sample underwent freezing, sectioning, and staining with hematoxylin and eosin (H&E). The inset displays an optical image of the H&E-stained tissue section from region 3, revealing a mixed histologic composition that includes both cancerous and adjacent normal stroma tissue [27]. Fig. 4(c) describes the pathological diagnosis of the five regions alongside the Lasso prediction outcomes.

### C. *Using Diffuse Reflectance Spectroscopy Gastrointestinal Cancers Using Diffuse Reflectance Spectroscopy*

Upper gastrointestinal tract cancers continue to be a major factor in the worldwide cancer burden. Precisely identifying tumor margins is essential for effective cancer removal and enhancing overall survival rates. Current real-time mapping techniques do not allow for a full resection margin assessment. Therefore, it is essential to evaluate the capacity of Diffuse Reflectance Spectroscopy (DRS) to distinguish between tissue types and offer real-time feedback to operators on gastric and esophageal cancer specimens.

Research has explored the creation of a DRS probe and tracking system, showing that machine learning can accurately differentiate between normal and cancerous tissue in real-time for gastric and esophageal samples [11]. Fig. 4(d) illustrates the DRS instrumentation used for ex vivo data collection. The DRS probe was linked to both the light source and the spectrometer to illuminate the sample and collect data. All electronic devices interfaced through custom software developed in Python, running on a laptop.

At each optical biopsy site real-time monitoring was performed including binary classification of each site on esophageal and gastric specimens, with the tissue type real-time displayed on the screen. Supervised machine learning and permutation feature importance were utilized for feature selection and performance evaluation of spectral data classification.

Results of the classification for stomach and esophagus spectral data were discussed using Light Gradient Boosting Machine (LGBM), Multilayer Perceptron (MLP), Support Vector Machine (SVM), and Extreme Gradient Boosting (XGB) classifiers. The XGB classifier demonstrated superior performance in differentiating between normal and cancerous tissues in both stomach and esophageal samples.

### D. *In vivo early lung cancer detection with Real-time endoscopic Raman spectroscopy*

Autofluorescence bronchoscopy (AFB) combined with white light bronchoscopy (WLB) is the most precise for localizing lung cancers within the central airways [28, 29]. However, the diagnostic accuracy of WLB+ AFB for detecting high-grade dysplasia (HGD) and carcinoma in situ can differ depending on the physician's level of experience. Raman spectroscopy, which investigates molecular vibrations to provide distinct spectral fingerprint-like features, offers high accuracy for classifying tissue pathology.

Fig. 4(a) shows the schematic diagram of the endoscopic laser Raman spectroscopy system. It illustrates the arrangement of the excitation and collection fibers. A diode laser generates the Raman excitation light, which is transmitted to the tissue surface using a removable fiber optic probe inserted through the instrument channel of the bronchoscope [30]. The same catheter collects emission from the tissue, which is then sent to the spectrometer for analysis. To correct spectral imaging distortion for better signal-to-noise ratio (SNR), The collection fibers are linked to the spectrograph via a unique round-to-parabolic fiber bundle.

One spectrum is captured per second, with clinical data gathered from 280 tissue sites in 80 patients [30]. Multivariate analysis and waveband selection techniques on the Raman spectra are used for detecting HGD and malignant lung lesions. For spectral classification, principal components with generalized discriminant analysis (PC-GDA) and partial least squares (PLS) are used [30]. Leave-one-out cross-validation (LOO-CV) is utilized, where each single spectrum is sequentially excluded for testing, and the remaining spectra are used for training. Waveband selection techniques such as stepwise multiple regression (STEP), genetic algorithm (GA), and LASSO enhance diagnostic specificity [30]. For 90% sensitivity, PLS analyses with STEP waveband selection provide the best specificity of 65% [30].

Combining Raman with WLB and AFB enhances the sensitivity of detecting HGD and malignant lesions compared to using WLB and AFB based diagnosis [30]. Through quantitative spectral analysis, the Raman algorithm holds promise to evolve into an automated and impartial diagnostic technique.

## IV. Different real-time Optical imaging based cancer diagnosis

This section discusses optical-based diagnosis, including real-time optical based diagnosis of colorectal cancer, optical diagnosis of neoplastic polyps during colonoscopy, and using pattern recognition and optical coherence tomography (PR-OCT) for real-time diagnosis of colorectal cancer..

### A. Colorectal cancer diagnosis based Real-time optical approach

The preresection accuracy of optical diagnosis of T1 colorectal cancer is discussed in [31]. It explains large non-pedunculated colorectal polyps (LNPCPs). Endoscopists, following a standardized procedure for optical assessment, predicted the histology of LNPCPs during colonoscopy in consecutive patients. They recorded the morphological features observed under white light, as well as the vascular and surface patterns identified using narrow-band imaging (NBI) [31]. A multivariable mixed-effects binary logistic LASSO model was employed to develop a risk score chart [31]. The reported sensitivity for optical diagnosis of T1 colorectal cancer was 78.7%, and the specificity for optical diagnosis of endoscopic unresectable lesions was 99.0% [31]. Notably, when endoscopists had low confidence compared to high confidence in their predictions, the optical diagnosis was often incorrect [31].

This suggests that considering an en bloc excision biopsy or consulting a trained colleague could be useful in further reducing the misidentification of T1 colorectal cancers. The ability to distinguish non-invasive LNPCPs from T1 colorectal cancers using a few white light and NBI features was quite effective [31].

### B. Pattern recognition Optical coherence tomography for Real-time colorectal cancer diagnosis

Optical coherence tomography (OCT) is a potentially alternative approach to endoscopic biopsy for differentiating normal colonic mucosa from neoplasia. A deep learning-based pattern recognition OCT system in [32] has been introduced to automate image processing. This deep learning based system provides accurate diagnosis, potentially in real-time. This automation helps with clinical translation, which is constrained by the need to process the large volume of generated data [32]. OCT, as an emerging imaging technique, obtains 3D "optical biopsies" of biological samples with high resolution. A convolutional neural network (CNN) has been employed to extract structural patterns from human colon OCT images [32]. The network was trained and tested using 26,000 OCT images gathered from 20 tumor areas, 16 benign areas, and 6 other abnormal areas [32]. Experimental diagnoses utilizing the pattern recognition (PR) OCT system was compared to established histologic findings, demonstrating a sensitivity of 100% and a specificity of 99.7%; the area under the receiver operating characteristic (ROC) curve was 0.998 [32].

Fig. 5(a) shows the swept-source OCT system. An illustration of RetinaNet is shown in Fig. 5(b), where the left portion features a Feature Pyramid Network (FPN) built on a ResNet-18 backbone, while the right portion includes two sub-networks that handle classification and localization predictions. The input light is divided by a 50-50 fiber coupler and then guided into a reference arm and a sample arm through two circulators. Two manual fiber polarization controllers are used to control fiber polarization. A variable density filter is employed to attenuate the reference arm. A galvo mirror system is used to scan the light beam in the sample arm. The interference signal, detected by a balanced detector (PDB), is transmitted to a data acquisition board [32]; then the monitor displays the real-time OCT B-scan images.

For pattern recognition the trained RetinaNet model was evaluated on the test cohort [32]. Given that the "Teeth" pattern signifies normal colon specimens, the network successfully identified all "Teeth" patterns within the test OCT images [32]. The PR-OCT workflow involved collecting colorectal B-scan images, separating them into training and testing sets, labeling "Teeth" and "Noise" patterns and feeding them into RetinaNet. Subsequently, the trained model was tested on the entire set of test images, followed by a performance evaluation.

Fig. 5(c)-(h) present pattern recognition outcomes from six representative OCT images [32]. In typical scenarios (Fig. 5(c)-(d)), "Teeth" patterns are identified and highlighted with green boxes, with the respective scores displayed next to each box [32]. Conversely, no such pattern is identified in the cancerous case (Fig. 5(e)). Fig. 5(f) illustrates the testing result for an adenomatous polyp, where no "Teeth" pattern was observed. In contrast, for treated complete responders, the "Teeth" patterns reemerge as depicted in Fig. 5(g). On the other hand, treated non-responders showed no such pattern detection (Fig. 5(h)).

### A. Optical Diagnosis of Neoplastic Polyps during Colonoscopy

Utilizing AI in real-time during colonoscopy assists colonoscopists in distinguishing between neoplastic polyps that need to be removed and non-neoplastic polyps that do not require removal. A multicenter clinical study compared a computer-aided diagnosis system with standard visual inspection of small (≤ 5mm in diameter) polyps in the sigmoid colon and rectum for optical diagnosis of neoplastic histology [33]. The computer-aided diagnosis system employs real-time ultra-magnification to visualize polyps during colonoscopy. After making a diagnosis whether neoplastic, uncertain, or non-neoplastic all polyps visualized through imaging were excised.

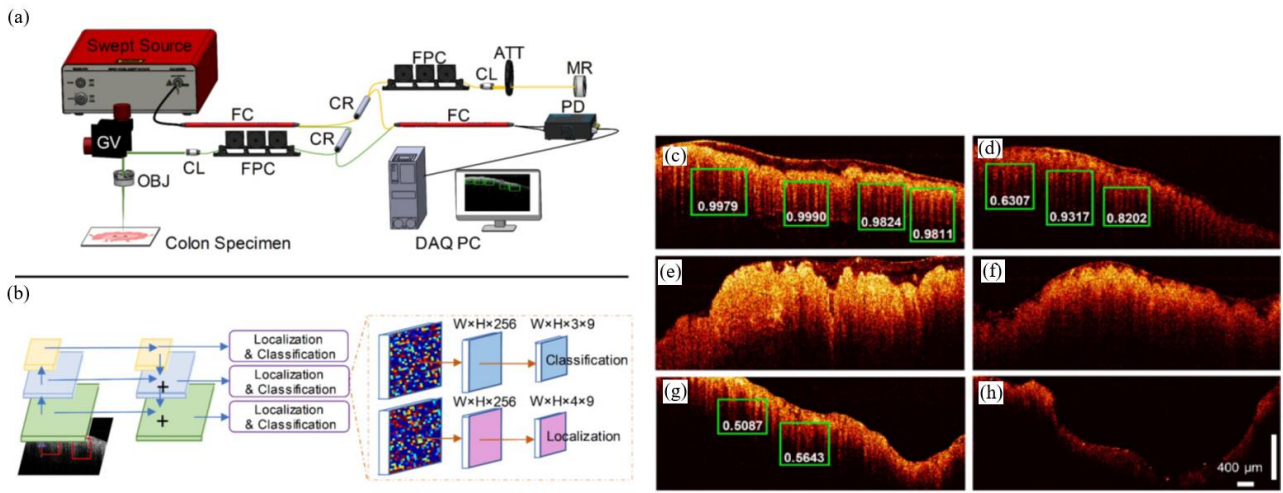

Fig. 5. Imaging framework of pattern recognition Optical coherence tomography (PR) OCT. (a) Homemade SS-OCT system, where FC, CR, FPC, CL, ATT, MR, GV, OBJ, PD, and DAQ PC stand for fiber coupler, circulator, fiber polarization controller, collimator attenuator, mirror, galvo mirror system, objective lens, photodetector, and data acquisition computer, respectively [32]. (b) RetinaNet framework through which an Feature Pyramid Network (FPN) with a ResNet-18 backbone (left part), and two sub-networks for predicting classifications and locations (right part) are shown.
pattern detection results based on PR-OCT for: (c)-(d) normal colon images (green boxes indicate predicted "Teeth" patterns and scores). (e) Cancer colon images. (f) Polyp colon images. (g) Treated complete responder colon images. (h) Treated non-responder colon images [32].

The study evaluated the sensitivity for neoplastic polyps achieved through computer-aided diagnosis and visual inspection, compared with histopathological examination. [33] suggests that using computer-aided diagnosis helped providers have higher confidence in optical diagnosis, potentially leading to cost reduction as more polyps could be left in situ. However, better confidence comes at a cost; computer-aided diagnosis assessment prolongs colonoscopy procedure time, which increases healthcare costs. It has been demonstrated that the time necessary for computer-aided diagnosis assessment of one small polyp is about 40 seconds. This is a trade-off, as computer-aided diagnosis reduces unnecessary removal of polyps and histopathologic assessment.

The sensitivity for neoplastic polyps was 88.4% with the standard method and 90.4% with the computer-aided diagnosis (CADx) method. The percentage of discordant pairs between the standard method and the CADx method was 7.2%.
The specificity for neoplastic polyps was 83.1% with the standard method and 85.9% with the CADx method. The discordance between the standard method and the CADx method was 7.9%.
The percentage of polyp assessments with high confidence for categorization into neoplastic or non-neoplastic polyp increased from 74.2% with the standard method to 92.6% with the CADx method. For small polyps during colonoscopy, AI polyp detection tools known as computer-aided polyp detection provide potential detection by up to 50%.

## V. ELASTOGRAPHY BASED CANCER DIAGNOSIS
This section describes different elastography procedures, starting with real-time shear-wave elastography. Subsequently, it elucidates the significance of real-time elastography (RTE)-guided biopsy in prostate cancer detection and diagnosis.

### A. Real-time Shear-Wave Elastography

In [34], two statistical analyses were performed to assess the effectiveness of real-time US shear-wave elastography (SWE) in identifying and pinpointing malignant prostatic lesions using both a patient-based and a region of interest (ROI) or sextant-based approach. The patient-centered approach accounts for the possibility of multiple findings within the same patient and their specific locations. For instance, a patient with a single false-positive lesion is managed in the same way as a patient with multiple false-positive lesions. This leads to variations in both the numerator and denominator of the false-positive rate compared to those in the sextant evaluation. Additionally, patients may present with both true-positive and false-positive findings. This accounts for the reduced diagnostic performance of SWE observed in the per-patient analysis [34]. The 95% confidence intervals for ROI-based estimates were consistently narrower than those for patient-based estimates in nearly all cases. In general, the sextant-based approach is more effective than the patient-based approach, as it allows for clinical treatment to be tailored to each individual sextant. The number of required biopsies can be decreased with a negative result in an SWE examination.
The role of RTE-targeted biopsy in the detection and diagnosis of prostate cancer is still challenging [34]. In [34] in order to assess the diagnostic accuracy of RTE-targeted biopsy applied the relative sensitivity value using the 10-core systematic biopsy as the reference standard. Based on this study it has been found that RTE-targeted biopsy did not outperform systematic biopsy [34]. In the core-by-core analysis, RTE-targeted biopsy identified a higher number of positive cores compared to systematic biopsy.

For some experiments it is shown that a favorable trend toward greater prostate cancer detection can be achieved while a combination of systematic biopsies and RTE-targeted biopsies was used. To do SWE acquisition the following steps are required: The elastographic box was maximized to cover half of the gland in a transverse plane. Each side of the prostate's peripheral zone was scanned separately from base to apex with a very slow motion to stabilize the signals. The two digital cine loops were saved for subsequent stiffness measurements. After completing the SWE acquisition, each digital cine loop was reviewed. Systematic stiffness measurements were conducted at the site of maximum stiffness within each sextant of the peripheral zone using a ROI shown in Fig. 6(a). It shows a SWE prostate measurement in a 58-year-old man with a PSA level of 4.6 ng/mL [34]. The top image displays the color-coded SWE within the color box. The bottom image illustrates the same plane with only B-mode information. Mean elasticity values in SWE were measured by placing two circular ROIs in the paramedian and lateral sextants [34]. The ROI was positioned on the peripheral zone using the B-mode image, where the peripheral zone was distinctly visible (shown in Fig. 6(b)). It shows a SWE prostate measurement in a 57-year-old man with a PSA level of 6.62 ng/mL [34]. The top image displays an orange-coded stiff area (solid-line ROI) incidentally found in the left base paramedian peripheral zone, which shows no abnormal features on the B-mode ultrasound (bottom image) [34].

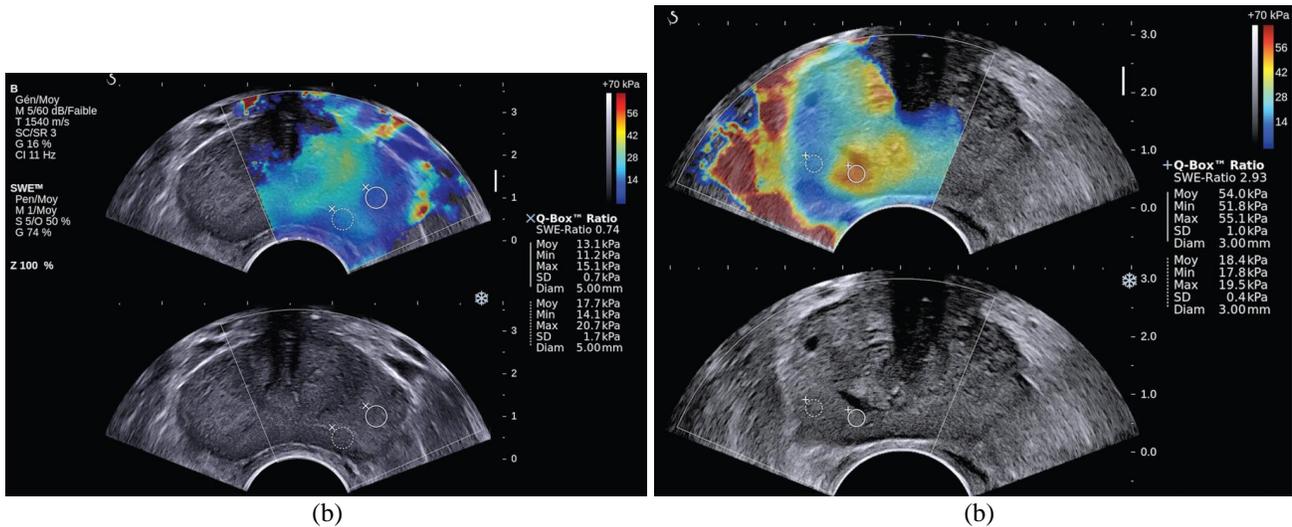

Fig. 6. (a) Shear wave elastography (SWE) prostate measurements in a 58-year-old male (with a prostate-specific antigen (PSA) level of 4.6 ng/mL). The upper image displays the color-coded SWE (with in the color box). In the lower image only B-mode data for the same plane is shown (b) SWE prostate measurements in a 57-year-old man (with a PSA level of 6.62 ng/mL). An orange-coded stiff area (incidentally found in the left base paramedian) is shown in the upper image while in the lower image, peripheral zone with no abnormal features on B-mode US is shown (from [34]).

The ROI, indicated by a dashed line, was situated in the lateral peripheral zone of the same sextant. The biopsy from the stiffest area revealed an 11-mm Gleason 7 adenocarcinoma [34].

### B. Video based Colorectal cancer invasion

Diagnosing the invasion depth of colorectal cancer using white light (WL) and image-enhanced endoscopy (IEE) techniques remains a significant challenge [35]. In order to determine the invasion depth of colorectal cancer, a dual-modal deep learning-based system, incorporating both WL and IEE, has been constructed and validated [35]. This approach combines features from both WL and IEE images for AI-based assessment of invasion depth in colorectal cancer [35].

WL and IEE images were paired together for analysis. In order to real-time evaluation of the endo-colorectal cancer system performance, thirty-five videos were used. Two deep learning models were developed: one using WL images (referred to as model W), and another using IEE images (referred to as model I). These models were compared with the endo-colorectal cancer system.

WL and IEE images are used in a CNN-based system for the diagnosis of colorectal cancer invasion depth [35]. This system was developed and evaluated across three hospitals, demonstrating its ability to detect lesions unsuitable for endoscopic resection in independent test images with high accuracy.

In Fig. 7(a), three models were developed for comparison the performance of WL images, IEE images, and the combination of WL and IEE images in diagnosing CRC invasion depth. These models are W, I, and Endo-CRC.

The WL image entered the WL convolution network, and the IEE image entered the IEE convolution network; then for feature concatenation a convolution block was used (Fig. 7(a)). Two pretrained CNN1 and CNN2 models are based on ResNet-50. These models were used to obtain eigenvectors from the WL and IEE images. In each image pair, the IEE vector was concatenated with the WL vector, and the resulting combined vector was used to generate feature maps. Finally, these feature maps were used to classify a self-constructed CNN3 [35]. To evaluate the clinical applicability of the endo-CRC system, video clips were used as the validation dataset.

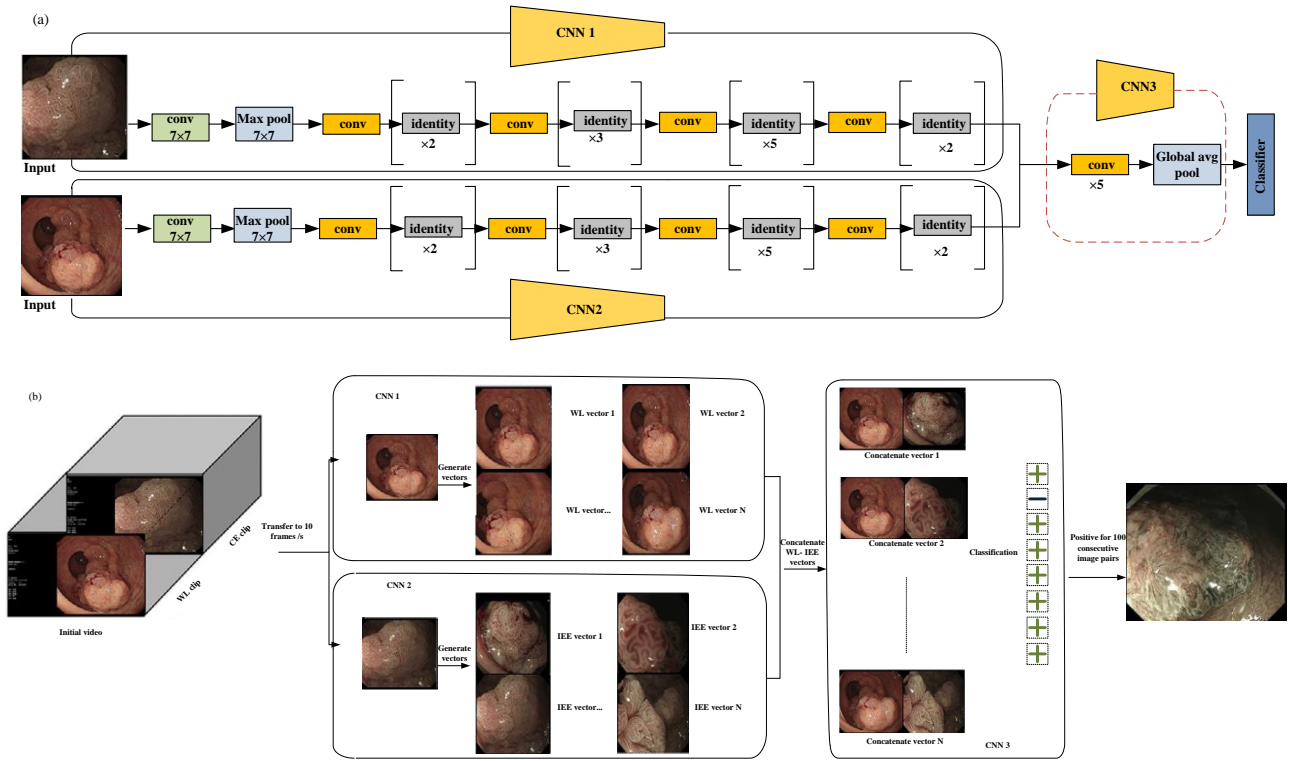

Fig. 7. (a) Multimodal data used for training in endo-colorectal cancer. White-light (WL) images and image-enhanced endoscopy (IEE) images are paired (with each pair containing one WL image and one IEE image). In order to obtain eigenvectors from the WL and IEE images, pretrained convolutional neural network (CNN) models (CNN1 and CNN2) based on ResNet-50) were employed respectively. Feature maps were generated from concatenated vectors (vector-IEE concatenated with vector-WL) to be used in classifying a self-constructed CNN3. (b) Workflow of video testing. with a rate of 10 frames per second, videos were converted to images. Each IEE vector was combined with every WL vector, sequentially. An unsuit-ER result was generated when 100 consecutive image pairs were classified as positive; otherwise, a suit-ER result was produced. 'Suit-ER' and 'unsuit-ER' stand for lesions suitable and unsuitable for endoscopic resection, respectively (adapted from [35]).

These videos were 10 to 19 seconds. As previously mentioned, the endo-CRC system diagnoses invasion depth using image pairs of lesions. Consequently, the endo-CRC system remains inactive and stores WL images at a rate of 10 frames per second (fps) during the WL clip [35].

When the video switches to IEE mode, each IEE image frame is paired with all WL images. The system then processes each image pair sequentially, outputting the results accordingly. A lesion was deemed unsuitable for endoscopic resection if 100 consecutive image pairs were classified as positive in each video. Otherwise, the lesion was considered suitable for endoscopic resection (suit-ER) (Fig. 7(b)). The system achieved a processing speed of 45 frames per second (fps) on the GPU. This processing speed fulfill the real-time processing speed requirement of 25 fps for endoscopic videos. Each video was converted to images at a rate of 10 frames per second. Subsequently, each IEE vector was sequentially combined with all WL vectors.

If 100 consecutive image pairs were classified as positive, the system determined the lesions to be unsuitable for endoscopic resection (unsuit-ER). Otherwise, the lesions were deemed suitable for endoscopic resection (suit-ER).

## VI. Cervical Cancer Detection using Neuromorphic

An ultra-fast, low-power dedicated neuromorphic hardware for intelligent classification to detect cervical cancer via utilizing single-cell and multi-cell images for has been introduced [36] This neuromorphic hardware employs the bio-inspired ASIC CM1K Neuromem chip [36]. The approach was evaluated using both single-cell and multi-cell images, with hardware performance benchmarked for accuracy, latency, and power consumption. These metrics were compared against traditional software implementations, including K-nearest neighbor (KNN) and support vector machine (SVM). Fig. 8 (a-d) illustrates the preprocessing steps applied to real dataset images. Preprocessing was carried out on an Intel i5 6th gen PC. The average preprocessing times were 335.5 ms for normal single-cell images and 136.3 ms for abnormal single-cell images. For multi-cell images, the average times were 728.6 ms for normal and 737.9 ms for abnormal images. Fig. 8(e)-(h) display the original image, top-hat filtered image, bottom-hat filtered image, subtracted image, binarized image, post noise-removal image, segmented image, and detected nucleus [36].

Fig. 8(i) depicts the complete process of learning, recognition, and classification within the diagnostic system. The nuclei area values were normalized to a range of 0 to 255 to ensure the system's independence from magnification. This normalization allowed the classification to be based on the relative sizes of the nuclei within the images. For each image used in training the database, one neuron was assigned, with the area metrics encoded into a byte array vector. This vector was then labeled with a category, either normal or abnormal, and loaded onto the neural network. The on-chip recognition was performed using the KNN mode. Testing vectors were broadcast for recognition one at a time. For single-cell images, the accuracy was comparable to, and in some cases surpassed, that of KNN and SVM classifiers. The Neuromem chip can be trained in two ways: by directly loading saved knowledge onto the chip (Save-and-Restore (SR) Mode) or by enabling the neurons to learn in real-time and build a knowledge base dynamically (Normal Mode).

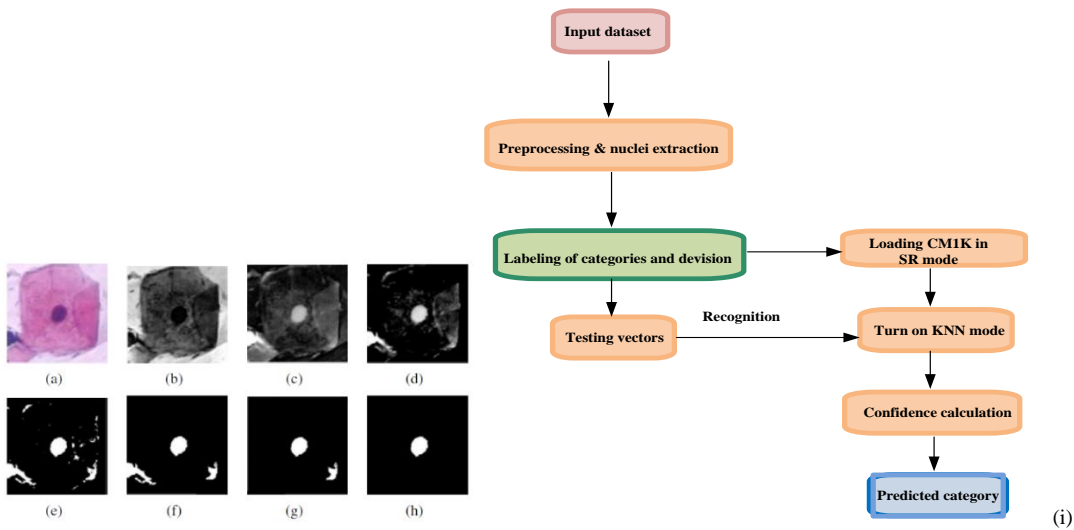

Fig. 8. Pre-processing steps and hardware processing flow: (a-h) show Original image, Top-Hat filtered, Bottom-Hat filtered, subtracted image, binarized image, post noise-removal, segmented image, and detected nucleus, respectively [36]. (i) Flow for the hardware accelerator technique containing learning, recognition, and classification. (from [36]).

In SR mode, training each neuron on the Neuromem chip took approximately 18 µs, compared to a few milliseconds for KNN and SVM classifiers. The hardware based recognition time /per vector doesn't rely on the size of the training dataset. In comparison to traditional KNN/SVM implementations, the average speedup in training time per vector is approximately 596 times for single-cell images and 477 times for multi-cell images. The average speedups in recognition time per vector for single-cell and multi-cell images are approximately 54 times and 41 times, respectively, compared to conventional methods.

This work is limited by its reliance on relatively basic preprocessing techniques and the use of a single feature for the 2-level classification of normal and abnormal samples. Enhancing the hardware approach with sophisticated preprocessing techniques, such as those achieved with deep learning [37] or GPUs, along with the utilization of multiple features [38], will significantly enhance the accuracy of the diagnostic system. Secondly, while the dataset for multi-cell images was small, the system's performance matching that of software implementations suggests that it could achieve similar results with larger and more varied datasets.

## VII. DIFFERENT FLUORESCENCE IMAGE BASED CANCER DIAGNOSIS

This section explains various fluorescence imaging-based methods for cancer diagnosis. It discusses a real-time fluorescence image guidance based on a goggle-augmented imaging as well as navigation system. Additionally, it covers breast cancer diagnosis through pH-based fluorescence lifetime endoscopy and examines the use of fluorescence lifetime imaging to delineate tumor margins in excised breast tissue samples.

### A. Goggle Augmented Imaging and Navigation System

In [39] for tumor resection a goggle augmented imaging and navigation system (GAINS) has been introduced. The system has shown high accuracy in detecting tumors in subcutaneous and metastatic mouse models. Human pilot studies involving breast cancer and melanoma patients have shown that GAINS can identify sentinel lymph nodes with perfect accuracy, achieving 100% sensitivity when using a near-infrared dye [39]. The clinical application of GAINS for guiding tumor resection and mapping sentinel lymph nodes holds significant potential for enhancing surgical outcomes. It aims to reduce the likelihood of repeat surgeries and improve the precision of cancer staging, thereby offering a more effective approach to cancer treatment.

Fig. 9(a) illustrates the design of GAINS. The processing unit generates co-registered composite color-fluorescence images, which are displayed in real-time via a lightweight and high-resolution head-mounted display (HMD) unit. The near-infrared (NIR) source is composed of light-emitting diodes (LEDs) paired with a bandpass filter. The display module features a high-resolution HMD. To align the imaging module with the camera's weight and the user's field of view (FOV), an adjustable and counterbalanced mechanical mounting system was integrated [39]. This system can improve user experience. As the concentrations of NIR contrast agents increase, GAINS determines the fluorescence intensity profile. To test GAINS functionality in vivo, a subcutaneous breast cancer mouse model was used. Through utilizing this image processing algorithm generates in real time, composite color-fluorescence images.

These images are concurrently displayed on both a HMD and a nearby personal computer. This system enables non-intrusive information display to the operating surgeon while simultaneously making the data available to the entire surgical team in the operating room. Overlaying fluorescence information on the normal visual field can facilitate rapid intraoperative tumor visualization. However, a larger sample size is necessary to confirm these results. Statistical analysis was conducted using OriginPro8. Paired t-tests were employed to compare fluorescence signals between tumors and background tissue in mouse models. The sensitivity of sentinel lymph node (SLN) detection was compared among the GAINS method, radioactivity, and blue dye techniques. These measurements showcase non-intrusive real-time image guidance. The minimal training needed has the potential to make this technology more accessible to clinicians.

### B. Breast cancer Diagnosis Using PH-Based Fluorescence Lifetime Endoscopy

A biopsy is often conducted during a surgical procedure for cancer diagnosis. Pathological biopsy of surgical specimens is also necessary to differentiate cancerous tissues from normal tissues. In [40], a novel method using fluorescence lifetime endoscopy (FLE) has been employed to discriminate between tumor and normal tissues (Fig. 9(b)). Reference [40] discusses the feasibility of real-time, in vivo, and in-situ cancer diagnosis by analyzing fluorescence lifetime data in tissues, eliminating the need for tissue resection. This study has been demonstrated in 20 mice, including two different types of mouse models with breast and skin cancer. The pH-related fluorescence lifetime results for normal and tumor tissues have been validated, showing consistency with the outcomes of the H&E staining test [40].

These findings are promising for intraoperative laparoscopic surgery applications, enabling assessment of cancer using an in vivo FLE system. However, for clinical applications, it will be necessary to perform pH assessment in vivo in tissues as an optical biopsy. Cancer diagnosis based on fluorescence lifetime has also been suggested for topical applications that reduce tissue absorption and fluorophore penetration. Fluorescein has been approved for clinical use through intravenous administration. In Fig. 9(b), the objective lens functions as an integral component, bridging the scan lens and the imaging fiber bundle system to form the endoscope system. The imaging fiber bundle system comprises a fiber bundle and an imaging lens, with the imaging bundle positioned at the focal point of the objective lens. The laser beam is directed through the scan lens into the imaging fiber bundle system to excite the tissue. The PMT which is connected to the digitizer, collected the reflected emission beam; this enables acquisition of 2D fluorescence lifetime information of tissues. In this system can be utilized in gastroenterology by inserting the imaging fiber bundle into the biopsy channel of a standard commercial endoscope. Fig. 9(c) shows images of the breast cancer model (top) and skin cancer model (bottom). The H&E staining results of tissues has been shown in Fig. 9(c), through which columns two and three display from the normal group and tumor group, respectively.

Tissues from the tumor group showed cell and nuclear damage compared to those from the normal group. The presence of tumor tissues over the evaluated regions of the tissue was validated using this technique. Cancer cells exhibited larger size, giant nuclei, dual nuclei, multiple nuclei, or irregular nuclei, and a different nucleus-to-cytoplasm ratio compared to normal cells [40].

Using fluorescence lifetime imaging for detecting Tumor margins in excised breast specimens in breast-conserving surgery, ensuring tumor-free surgical margins is crucial. However, in as many as 38% of cases, a second surgery becomes necessary because malignant cs are detected at the edges of the excised resection specimen.

Consequently, sophisticated imaging tools are essential to guarantee clear margins during surgical procedures. A random forest classifier has been investigated to diagnose tumors at resection margins. This provides an intuitive visualization of probabilistic classification on tissue specimens.

Fluorescence lifetime imaging (FLIm) measurements is used in this classifier. FLIm uses parameters derived from point-scanning label-free of breast specimens. FLIm data obtained from fresh lumpectomy and mastectomy specimens were used [41]. A registration technique between autofluorescence imaging data and cross-sectional histology slides is used to conduct the supervised training [41]. This approach offers high prediction accuracy, swift acquisition speed for large area scans, spatial refinement for suspicious regions, and nondestructive probing [41]. This method offers intuitive visualization of tissue characteristics and achieves high tumor classification sensitivity and specificity of (89%) and (93%) respectively.

Figure 10(a) displays a prototype time-domain multispectral time-resolved fluorescence spectroscopy (ms-TRFS) system featuring an integrated aiming beam [41]. During imaging the fiber probe was manually guided (Fig. 10(b)) for all specimens.

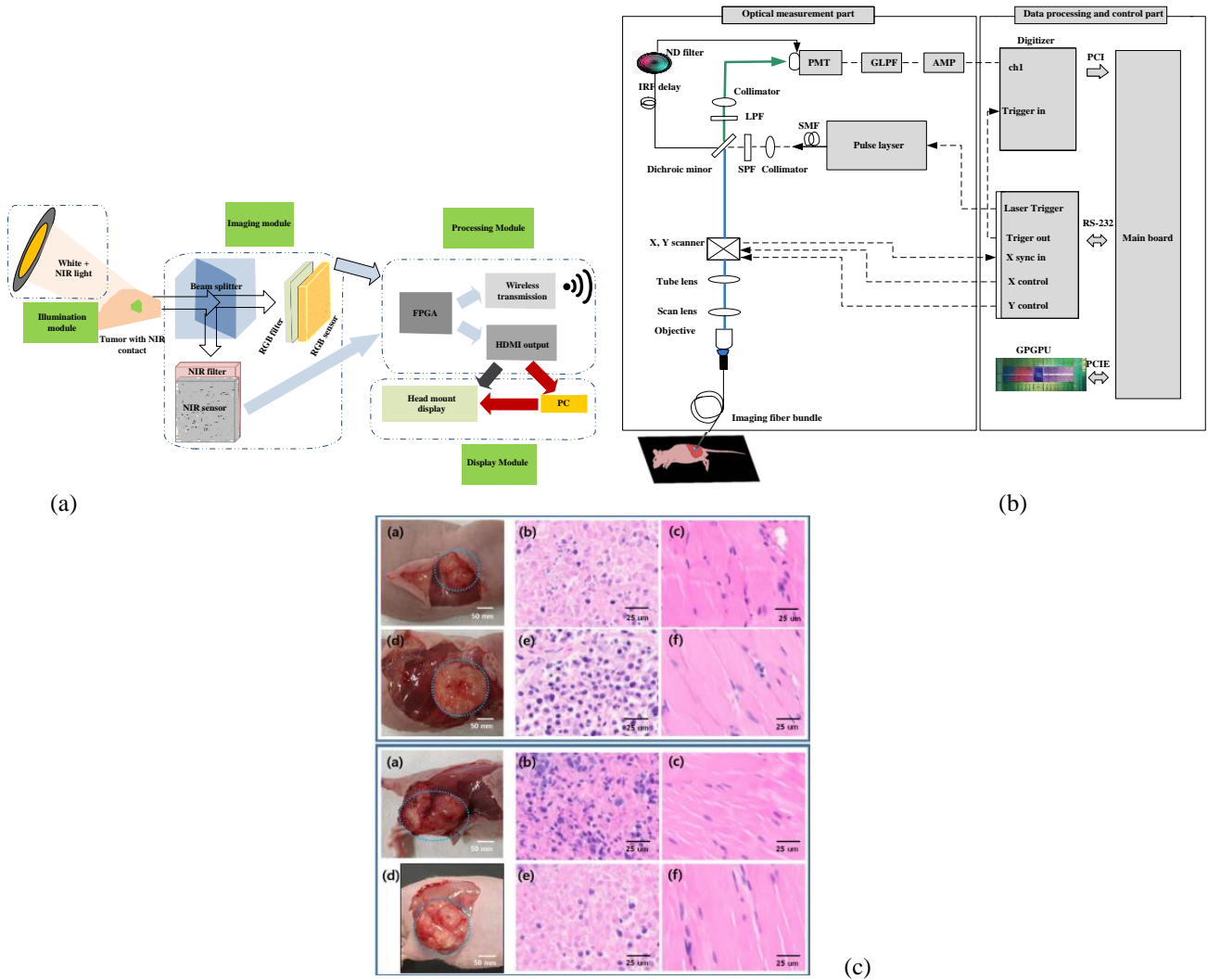

(a)

(b)

(c)

Fig. 9 (a) Framework of the goggle augmented imaging and navigation system (GAINS) system with its different modules and information flow between the modules. ( [39]). (b) Schematic of the experimental set-up for fluorescence lifetime endoscopy (FLE) [40]. (c) Images at the top show the breast cancer model, and those at the bottom show the skin cancer model. (b) H&Estaining results of tissues from the normal group, and (c) from the tumor group (adapted from [40]).

Throughout the scanning process, During the scanning process, the distance between the sample and the probe was maintained at a few millimeters. When the fiber tip strayed too far from the sample, the computer detected a drop in signal amplitude and triggered an acoustic alarm to alert the operator to reposition closer to the sample. An aiming beam acting as a marker to superimpose fluorescence data onto the video is integrated into the optical path. The FLIm system and associated computers are mounted on a cart, with dual screens are shown in Fig. 10(c) to image the breast specimens.

Fig. 10(d) shows the training pipeline. A semiautomatic histology registration procedure was used to map the histology slide and the pathologist's annotations onto the video image [41]. Pathologist tracings from the histology are transferred to the video domain. Fluorescence parameters from the resulting regions are then sent to a random forest classifier.

The output color is used to encode the type of tissue in color and saturation represents the certainty of the probabilistic output. Fig. 10(f) displays the overall accuracy which is varying the percentage of training samples in 2% increments, ranging from 2% to 100%. When over 75% of the training samples are utilized, the results stabilize, achieving an overall accuracy of approximately 97.5%.

### D. Digital biopsy

Microscopic analysis of tissue is the definitive method for detecting cancer [42]. Conventionally, the reporting of prostate biopsies using hematoxylin and eosin (H&E) staining involves fixation, processing, the preparation of glass slides, and analysis using an analog microscope by a local pathologist [42]. Real-time remote access and image digitalization are employed to streamline the reporting process and provide a foundation for AI and machine learning.

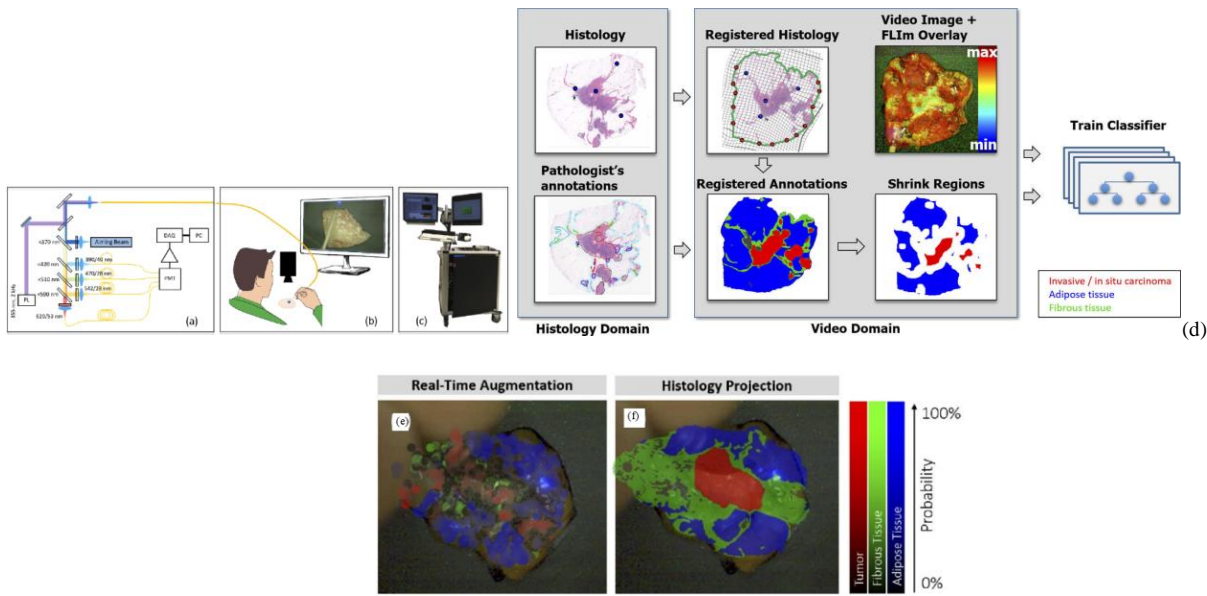

Fig.10 (a) the ms-TRFS imaging framework. For excitation and autofluorescence collection, a single fiber is used. PL, DAQ, and PMT represent pulsed laser, data acquisition, and photomultiplier, respectively [41]. (b) Imaging setup is done as follows: A manual scan was conducted for each specimen. In order to overlay fluorescence data onto the video at the measurement site, an aiming beam embedded in the optical path was used as a marker. (c) The FLIm system and computers were configured into a setup (that included two screens) to image the breast specimens, [41]. (d) A supervised pipeline utilizes used for training in which for registration it uses cross-sectional histology and the video image. Within this method the pathologist tracings translated from the histology to the video domain, and a random forest classifier is used for classification, fed by fluorescence parameters from the resulting regions. (e), (f) For invasive carcinoma, augmented overlay with histological ground truth reveals a marked decrease in prediction accuracy., with an erroneous classification of tumor through the overlay [41].

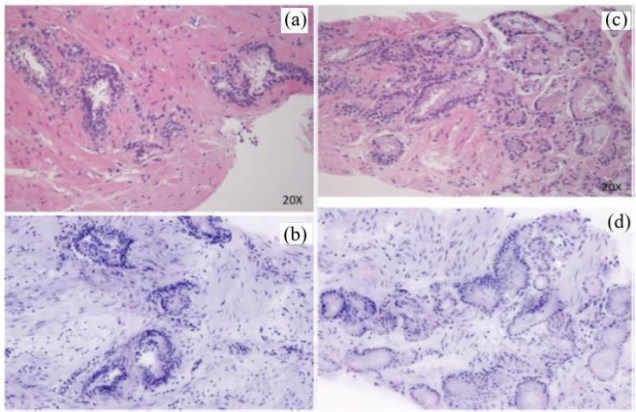

Fig. 11. Noncancerous prostate tissue in (a) HE and (b) FCM. Cancerous prostate glands in (c) HE and (d) FCM. FCM and HE stand for fluorescence confocal microscopy and hematoxylin-eosin, respectively (adapted from [42]).

A new optical technology named fluorescence confocal microscopy (FCM) allows immediate digital image acquisition similar to H&E staining, eliminating the need for conventional processing methods. Prostate biopsies were conducted at a single coordinating unit to assess consecutive patients (based on clinical indications) [42]. FCM digital images obtained immediately from prostate biopsies were stored; then, after applying conventional H&E processing on the glass slides, they were digitized and stored.

Fig. 11 shows noncancerous prostate tissue in (a) H&E and (b) FCM, and cancerous prostate glands in (c) H&E and (d) FCM. The minimum sample size was determined based on the accuracy of sensitivity or specificity estimates in diagnostic test studies. This calculation ensures that the sample size is sufficient for estimating inter-rater agreement using Cohen's kappa, with specified confidence limits reflecting the accuracy of sensitivity or specificity estimates. FCM with VivaScope has the potential to elevate microscopic analysis in real-time. The VivaScope is compact and space-efficient, making it easily adaptable for use in the operating room. Moreover, acquiring digital images is straightforward and does not necessitate specific technical skills.

## VIII. HYPERSPECTRAL IMAGING BASED CANCER DIAGNOSIS

In this section, different hyperspectral imaging techniques will be discussed. First, VNIR–NIR hyperspectral image fusion for brain cancer detection and in vivo skin cancer assess via millimeter-wave based imaging are discussed. Next, for brain cancer detection during surgical operations, using hyperspectral images with spatio-spectral classification are discussed. After that hyperspectral imaging based intraoperative visualization system are discussed. Finally, in vivo identification of glioblastoma tumors based on hyperspectral images and employing a deep learning framework is discussed.

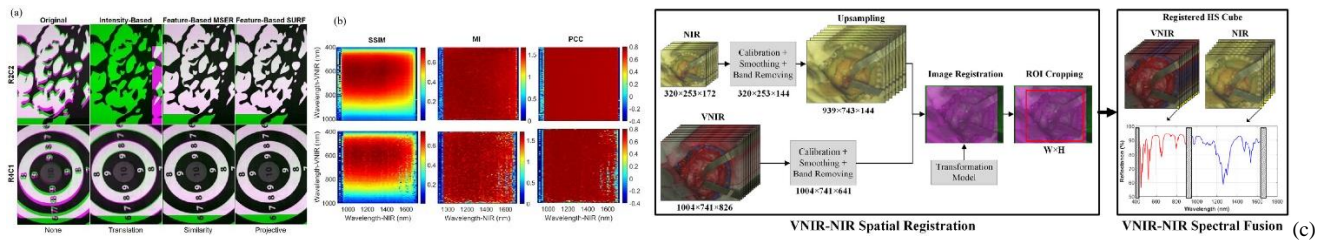

Fig. 12. (a) Spatial registration process of VNIR–NIR [43]. Different registration techniques are applied to two registration result examples (R2C2 and R4C1). For image overlap visualization VNIR is in green and NIR is in magenta. The default registration is shown in the first column without applying any transformation to the data. The second column shows intensity-based results, the third column shows feature-based with MSER, and feature-based with SURF techniques are shown, in the fourth columns. (b) In order to identify the suitable spectral bands for registration (via using the feature-based SURF technique with projective transformation) is performed with coarse search results of the structural similarity index measure (SSIM), mutual information (MI), and Pearson's correlation coefficient (PCC) [43]. (c) VNIR–NIR spatial registration combined with spectral fusion flow diagram of the processing framework (adapted from [43]).

### A. Image fusion based on VNIR–NIR hyperspectral images for detection of Brain cancer

For brain tumor resection assistance during surgery it is observed that intraoperative guidance tools have limitations [43]. During surgery in real time, in order to delineate brain tumor tissue, hyperspectral imaging offers new advancements. However, hyperspectral acquisition systems face limitations in spatial and spectral resolution. To improve spatial and spectral resolution, image fusion combines information from different sensors.

In [43] an intraoperative hyperspectral acquisition system for image fusion uses two push-broom hyperspectral cameras covering the visual and near-infrared (VNIR) [400–1000 nm] and near-infrared (NIR) [900–1700 nm] spectral ranges. In order to obtain a hyperspectral cube with a wide spectral range [435–1638 nm] both hyperspectral images were registered using intensity-based and feature-based techniques. For the hyperspectral imaging registration dataset, R2C2 and R4C1 are provided as two examples, as shown in Fig. 12(a). The initial column presents the registration result without any geometric transformation. The following columns display the optimal outcomes from each registration method and their corresponding best geometric transformations. Green-magenta false-color images overlay the VNIR and NIR pseudo-RGB images. Misregistration between the VNIR and NIR images are shown with Magenta and green pixels, while gray-scale pixels indicate similar intensity values in the two registered images.

In the intensity-based registration employing the translation transformation, R2C2 is incorrectly registered, and compared to the outcome without employing any transformation, R4C1 improves the registration. The reason for these incorrect registrations, is the random noise that is found in some spectral bands. This random noise affects the maximum intensity. The feature-based maximally stable extremal regions (MSER) approach, combined with similarity transformation, enhances the effectiveness compared to traditional intensity-based methods. though in both images some misregistered pixels can be observed. Using projective transformation, the feature-based speeded up robust features (SURF) technique had the best results. A coarse-to-fine search was applied to identify the VNIR and NIR bands. This was achieved using grayscale images from a single spectral band captured by both cameras.

Fig. 12(b) shows the R2C2 and R4C1 heatmaps resulting from the coarse search using the similarity index measure (SSIM), mutual information (MI), and Pearson's correlation coefficient (PCC) metrics [43]. Due to considering image structure, the SSIM metric provides the highest results, showing the regions 500–700 nm and 950–1500 nm in the VNIR and NIR ranges, respectively [43]. The other metrics only consider image intensity.

In the intensity-based registration employing the translation transformation, R2C2 is misaligned. Compared to the result without using segmentation maps obtained with the K-means algorithm R4C1 improves the registration [43]. For every hyperspectral image, the Jaccard metric was calculated based on the ground-truth image and the segmentation map. The VNIR data exhibited superior performance in color segmentation when analyzed with the K-means algorithm. This was followed by the fused data, which showed improved segmentation results using the K-medoids and hierarchical K-means (HKM) algorithms. Material identification performed exceptionally well using NIR data across all three algorithms. The application of HKM to NIR data led to significant improvements in material-color segmentation, while additional enhancements were observed in the fused data processed with K-means.

Statistical analysis of the segmentation results was conducted using a paired, one-tailed Student's t-test with a significance level of 5%. The proposed approach in [43] uses a two-step method which is composed of VNIR–NIR spatial registration and VNIR–NIR spectral fusion (Fig. 12(c)).

The raw VNIR and NIR images undergo image calibration to counteract environmental illumination effects, noise filtering, and band removal processing steps. Then in order to enable image registration, the VNIR pixel size has been matched the NIR image via up sampling. This up sampling is achieved via employing a generated transformation model, in which the fixed image is the VNIR, and the moving image is the NIR.

In order to obtain the same region of interest (ROI), after registration of VNIR and NIR images, they are cropped. Finally, in the last stage, the spectra from both VNIR and NIR images are combined. In order to facilitate spectral fusion a reflectance offset has been applied to the NIR spectrum.

Through VNIR–NIR data fusion, compared to using each data modality independently, the classification results are improved by up to 21% in accuracy.

### B. Detection of in-vivo skin cancer with Millimeter-wave imaging

High-resolution millimeter-wave imaging (HR-MMWI) has the potential to provide affordable noninvasive tissue diagnostic information due to its ample depth of penetration and significant differentiation [44]. An HR-MMWI system was utilized to evaluate its application in in-vivo skin cancer diagnosis. The assessment involved measuring benign and malignant skin lesions from 71 patients with various conditions such as melanoma, basal cell carcinoma, squamous cell carcinoma, actinic keratosis, melanocytic nevi, angiokeratoma, dermatofibroma, solar lentigo, and seborrheic keratosis [44].

The downscaling of sensors and antenna interfaces can be performed which it stems from the short wavelength and hyperspectral characteristics of millimeter waves, making them ideal for the development of handheld or point-of-care devices. As a result, a handheld configuration can be designed instead of a benchtop HR-MMWI system, offering in-vivo imaging of skin tissue at a remarkably low production cost. This will help dermatologists and dermatologic surgeons acquire images of cancer tissues that resemble histopathological analysis. This can be performed either before a biopsy or after tumor excision, enabling the prompt identification and removal of any remaining tumor tissue.

Priced similarly to a dermatoscope, this portable, real-time imaging tool allows dermatologists to efficiently see more patients in multiple rooms. By utilizing MMWI, the reduction in unnecessary biopsies can exceed 50%. HR-MMWI requires less training and expertise from users. This rapid classification of lesions within 20 seconds represents a groundbreaking clinical advancement in the detection and management of skin cancer.

Fig. 13 depicts the newly developed ultra-wideband millimeter-wave imaging system, which operates within a bandwidth of 98 GHz (12–110 GHz) [44]. During each scanning step, two sub-band antennas are positioned in front of the target to transmit signals within their respective subband frequency ranges and capture the backscattered signals.

The workflow involves image reconstruction, processing, depth estimation, 3D PCA-based feature extraction, feature classification, and cancer detection based on malignancy scores. Initially, a multivariate statistical analysis based on PCA is employed to automatically identify the most diagnostically significant features.

PCA condenses nearly 1000 intensity variables from raw tissue images into six principal components (PCs), each representing a linear combination of the original variables. The number of PCs is determined from the millimeter-wave images using singular value decomposition (SVD) [44]. Once the PCs are chosen, the scores and loadings are aggregated across all points and all PCs.

The PC scores were collected and used as input for the classifier model. Various linear and nonlinear classifiers, including Linear Discriminant Analysis (LDA), K-Nearest Neighbors (KNN) with different K-values, Linear Support Vector Machine (LSVM), Gaussian SVM (GSVM) with different margin factors, and Multilayer Perceptron (MLP), were tested using all possible combinations of features [44].

The classifiers were selected based on the dataset size to ensure high reliability and robustness in classifying lesion types.

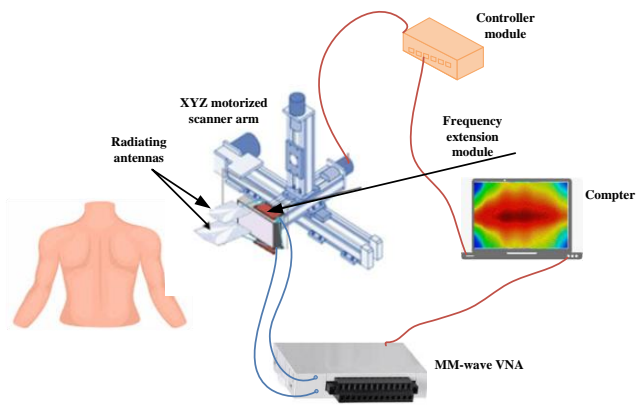

Fig. 13. A framework for of an ultra-wideband millimeter-wave imaging. With this system a synthetic bandwidth of 98 GHz can be achieved. In order to transmit and record backscattered signals, two sub-band antennas are positioned in front of the target at each scanning step ([44]).

### C. Detection of Brain Cancer During Surgical Operations Using Spatio-Spectral Classification of Hyperspectral Images

Treating brain cancer surgically poses a substantial challenge in neurosurgery [45]. Tumors often infiltrate diffusely into the surrounding normal brain, complicating their accurate identification visually. Surgery remains the primary treatment for brain cancer, and achieving precise and complete tumor removal can enhance patient survival rates.

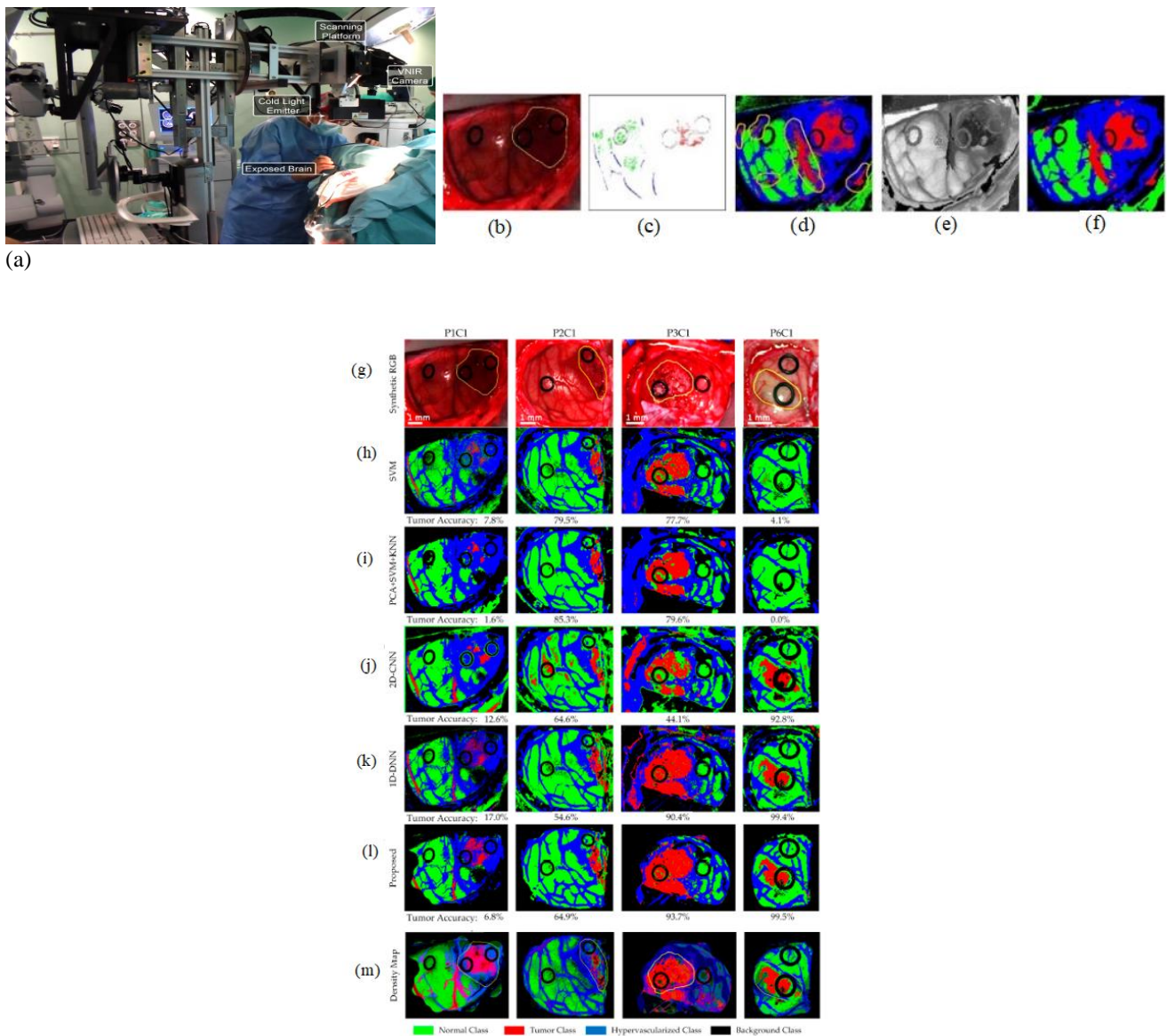

Fig. 14 (a) A hyperspectral acquisition system as an intra-operative system (utilized during a neurosurgical procedure). An optimized spatial-spectral supervised classification algorithm is used which its results in each step are as follows: (b) Synthetic RGB images generated from the hyperspectral cubes. (c) the supervised classification training with golden standard maps. (d) Supervised classification maps after applying the SVM algorithm. (e) the hyperspectral cubes' FR-t-SNE single-band visualization (f) Classification maps optimized spatially after using the KNN filter. (from [45]). Four test hyperspectral images classification maps with tumor accuracy mentioned below each map. (g) Highlighted area in yellow shows the synthetic RGB image with tumor (h–l) The results of multiclass classification maps via using SVM, PCA + SVM + KNN, 2D convolutional neural network (2D-CNN), one-dimensional 2D deep neural network (1D-DNN), and the proposed framework, respectively [47]. Green shows normal tissue, red shows tumor tissue, and hypervascularized tissue is shown in blue; black indicates the background. (m) Density maps resulting from the surgical aid visualization algorithm. The probability values (determined with the majority voting algorithm) are shown through colors. (adapted from [47]).

Hyperspectral imaging is a noncontact, non-ionizing, and non-invasive tool used during surgery to visualize brain tumors in real-time, assisting neurosurgeons in tumor resection. [45] presents a classification method that utilizes both spatial and spectral features within hyperspectral images. This method aids neurosurgeons in precisely identifying tumor boundaries during surgical procedures.

Fig. 14(a) shows the intra-operative hyperspectral acquisition system. The classification framework comprises five main stages: data pre-processing, dimensionality reduction, spatial-spectral supervised classification, unsupervised clustering segmentation, and hybrid classification [45]. Fig. 14(b-f) presents the outcomes of each phase in the optimized spatial-spectral supervised classification of the patients [45].

Fig. 14(b) shows synthetic RGB images generated from the hyperspectral cubes, and Fig. 14(c) depicts the reference standard maps employed for supervised classification training. Fig. 14(e) presents the FR-t-SNE representation of the hyperspectral cube using a single band [45]. These images highlight distinct areas on the brain surface, making it easier to identify their borders, including the tumor area.

The FR-t-SNE outcomes, combined with probability scores from the supervised classification maps, are inputted into the KNN filter. After the KNN filtering process, the spatially optimized classification map is obtained, as shown in Fig. 14(f). The regions belonging to each class in the images have been made more uniform, resulting in coherent classification maps. While the distinctions between the supervised classification maps and the spatially optimized classification maps may be subtle to the naked eye (Fig. 14(d) and Fig. 14(f)), this homogenization is an important process that enhances the final phase of the cancer detection algorithm [45].

During the spatial-spectral supervised classification stage, a comparison was made between the sequential time results from a CPU implementation and the time results achieved using a hardware accelerator [45]. The accelerated version of the algorithm requires transmission time due to the connection between the computer and the hardware accelerator. When the hardware accelerator is used in the spatial-spectral supervised classification stage, results in an average speedup factor of 26.83x. During surgery with utilized system in [45], it takes approximately 1 minute to produce a classification map of the captured scene. Some false positives have been absorbed in the results; additionally, Validation of the system in clinical settings is necessary, as it has shown some misclassifications among different types of tissues., and the entire algorithm needs to be accelerated.

### D. Enhanced Brain Tumor Delineation Using an Intraoperative Visualization System with Hyperspectral Imaging

Hyperspectral imaging allows for the capture of a broad spectrum of wavelengths across the electromagnetic spectrum from the surfaces of scenes observed by sensors. By leveraging this data along with a range of classification algorithms, it is possible to accurately identify specific materials or substances present in each pixel. [46] during neurosurgical operations (in order to distinguish tumor tissue from brain tissue) exploits the characteristics of hyperspectral imaging to develop a prototype which can enhance the precise outlining of tumor boundaries and improve the results of surgery.

For real-time processing, during surgical procedures a hardware accelerator is utilized in conjunction with a control unit [46]. This hardware accelerator expedite the hyperspectral brain cancer detection algorithm. A labeled dataset is used as the training data for the supervised phase, demonstrating that the system can distinguish between normal and tumor tissue in the brain during in vivo analysis.

The system provides results within about one minute during surgery, presenting a practical tool that could enhance excision and potentially enhance patient outcomes. The acquisition system incorporates a software platform that has been created by integrating three distinct software development kits [46]. These kits are from two types of hyperspectral cameras and a stepper motor controller [46]. The VNIR initiates the capture process from right to left across the platform equipped with the stepper motor. After capturing with VNIR, the stepper motor halts at the final position to stabilize the system for a few milliseconds before setting the speed. Subsequently, NIR begins capturing from left to right on the platform. The stepper motor repositions the scanning platform to the central position. This technique accelerates the acquisition process by 3 times compared to the original software. The control unit manages preprocessing, HKM clustering, and the majority voting algorithm. Simultaneously, the hardware accelerator oversees the spatial-spectral supervised classification stage, incorporating PCA and SVM algorithms.

Due to high computational demands for processing of thematic maps (which are from a validation database of hyperspectral images) it is carried out by the hardware accelerator [46]. Both the unsupervised and supervised stages run concurrently and then the control unit runs the majority voting algorithm to produce the final HELICoiD three maximum density (TMD) map [46]. The brain tumor is marked in red in the TMD map. This creates an RGB representation based on the top three major probabilities per cluster derived from the HKM clustering algorithm. This image is shown to the neurosurgeon through the hyperspectral processing interface.

### D. Deep Learning-Based Framework for In Vivo Identification of Glioblastoma Tumor using Hyperspectral Images of Human Brain

To ensure accurate guidance for tumor removal in real-time during neurosurgery, it is essential to develop a method that does not rely on labeling. Hyperspectral imaging can aid surgeons without the need for any contrast agent. In [47] proposes a framework to process in vivo human brain tissue with hyperspectral images which is based on deep neural networks. A human image database is used for the evaluation of the framework. This framework produces a thematic map that outlines the brain's parenchymal area and pinpoints the tumor's location. A pipeline is used for data processing. This pipeline provides a density map in which gradient colors are used to represent normal, tumor, and hyper vascularized tissues [47]. The results obtained using the proposed framework in Fig. 14(g) are promising, particularly for Patient 6, where identifying the tumor location with the naked eye was extremely challenging.

A synthetic RGB image is shown in Fig. 14(g). In this image, yellow lines are utilized to outline the tumor area. Figs. 14(h)–(l) display the results of the multiclass classification maps using SVM, PCA + SVM + KNN, 2D convolutional neural network (2D-CNN), one-dimensional 2D deep neural network (1D-DNN), and the proposed framework in [47], respectively. Normal tissue, tumor tissue, and hyper vascularized tissue are represented by green, red, and blue colors, respectively, while black indicates the background. Fig. 14(m) showcases the density maps generated by the surgical aid visualization algorithm, which are derived from the optimal threshold set for the tumor class.

The colors in these density maps have been modified according to the probability values obtained from the majority voting algorithm. The deep learning pipeline achieved an overall multiclass classification accuracy of 80%, surpassing the performance of traditional SVM-based methods. The proposed framework [47] employs a DNN for classification due to its shorter execution time. The CNN takes about one minute to process each hyperspectral cube, significantly longer than the DNN, which completes the task in approximately 10 seconds per cube. The training dataset had significantly fewer samples in the tumor class compared to the other classes, which led to an equal number of samples across all classes. This balance effectively minimized the overall training duration [47]. However, the use of real-time image fusion has some limitations, such as the substantial expenses associated with these systems, along with the extra time needed from physicians for examinations and image fusion.

## IX - PET/MRI IMAGE DEEP LEARNING BREAST CANCER

For advanced breast cancer in order to determine patient responses to neoadjuvant chemotherapy (NAC), CNNs with PET and MRI images as input have been used [48]. CNNs are employed for data classification, and AlexNet, a CNN variant, optimizes computational efficiency and boosts accuracy through the integration of dual convolution layers, enabling a comprehensive evaluation of the response to NAC. In order to determine the presence or absence of a complete chemotherapy response, it has been shown that CNNs improve this. The dataset used in the study was limited in size. CNNs, while capable of analyzing complex image features, typically necessitate a larger dataset to achieve optimal performance. Therefore, to overcome this limitation employing K-fold validation is important. The dataset exhibited a significant imbalance between responders and non-responders, potentially leading to an overestimation of accuracy. Such imbalance may yield results that are highly misleading. Additionally, unlike the conventional method, the CNN method needs a larger sample population since it did not incorporate changes between the baseline and interim images.

## X. FUTURE WORK

In this paper, we aim to review and summarize the latest literature on real-time AI based cancer diagnosis. The review is centered on different imaging methods as well as image fusion techniques. We discuss the pros and cons of the approaches reported in the literature. We summarize these challenges as follows in Table I.

## XI. CONCLUSION

This paper covers various aspects of real-time AI-based cancer diagnosis, including image fusion, spectroscopic techniques, optical imaging, elastography, neuromorphic methods, fluorescence imaging, and hyperspectral imaging. Key findings include:

### 1. ULTRASOUND IMAGE FUSION:

-Computer-aided surgical navigation systems that are equipped with image-fusion and image-overlaying capabilities can enhance the digitally-coordinated precision of minimally invasive urology; since this systems are integrated with real-time organ tracking and robotic systems. These advancements also improve predictive abilities, guiding surgeons toward optimal surgical outcomes.
-A fusion system combining transrectal ultrasound (TRUS) with magnetic resonance imaging (MRI) could address the limitations of relying solely on 2D grayscale TRUS for intraoperative guidance.

-The fusion of endoscopic ultrasound (EUS) and computed tomography (CT) images allows for easier navigation and profiling of target tumors and adjacent anatomical structures. This integration method can significantly shorten the learning process for mastering and navigating EUS procedures.

-Incorporating ultrasound (US) into image fusion offers several benefits, including real-time images for image-guided interventions, the absence of radiation exposure for both patients and staff, and the ability to compare findings between different modalities.

-The advantages of real-time image fusion in liver include improved visualization and targeting during therapeutic procedures, leading to reduced radiation exposure for both medical staff and patients.

-In prostate imaging electromagnetic needle tracking is both feasible and allows for off-plane biopsy and intervention in lesions containing air.

- Augmented reality navigation has the potential to significantly improve decision-making during percutaneous interventions, endoscopic surgeries, and extracorporeal therapies.

### 2. DIFFERENT SPECTROSCOPY TECHNICS FOR REAL-TIME IN VIVO CANCER DIAGNOSIS

- Machine learning based real-time distinguish between gastric and esophageal cancer and healthy tissue has been done via ex vivo validation of the diffuse reflectance spectroscopy (DRS). This marks a significant advancement toward creating a real-time, in vivo tool for tumor mapping. In order to evaluate the DRS real-life utility for resection tumor margin and the their ability in improving long-term outcomes, external validation and intraoperative testing are needed.
- In order to thoroughly assess the effectiveness of Raman spectroscopy technology in improving the localization of lung cancer and identifying precancerous lesions, Additional multi-center clinical trials are necessary.

- By combining with highly sensitive diagnostic optical imaging methods such as fluorescence endoscopy, the Raman technique is becoming a powerful diagnostic tool for routine bedside evaluations and intraoperative procedures.

-In order to prove that RTE-targeted biopsy is superior to systematic biopsy, there are inadequate reasons; However, combining both techniques enhances the prostate cancer detection accuracy.

### 3. DIFFERENT REAL-TIME OPTICAL IMAGING BASED CANCER DIAGNOSIS

-The number of patients currently benefiting from optical diagnosis is limited. In order to provide the absolute risk for T1 colorectal cancers a validated score chart has been developed. This chart uses real-time information on the presence or absence of white light and narrow-band imaging (NBI) features, which could be an important step in defining strict indications for endoscopic submucosal dissection only for lesions with a high risk of T1 colorectal cancer. Future study is need to determine how the risk score chart can be improved needs to be ultimately used for clinical decision-making. Through this clinical decision making it must be clarified the type of endoscopic resection and whether to proceed to surgery instead of endoscopy.

- The network must be trained on a larger number of patients. Assessing the novel "wait and watch" rectal cancer treatment management strategy could be a potential future improvement. This approach enables treatment responders with no residual cancer to be safely monitored through imaging instead of surgery, thereby preserving their quality of life.

### 4. ELASTOGRAPHY BASED CANCER DIAGNOSIS

-Shear wave elastography (SWE) assess the tissue stiffness, offering supplementary information for detecting and guiding biopsies in prostate cancer. SWE exhibits high sensitivity for detecting prostate cancer and has a very high negative predictive value (NPV), minimizing the likelihood of missing any cancers or only a few in the peripheral zone of the prostate. Elasticity values of prostate cancer rise with the Gleason score, indicating a correlation with the aggressiveness of prostate cancer, particularly in cases with elevated Gleason scores. Clinical rezults indicate that real-time elastography (RTE)-targeted biopsy did not surpass systematic biopsy in overall prostate cancer detection or in overall prostate cancer detection, including initial biopsies for men suspected of having the disease. Nevertheless, a core-by-core analysis revealed that RTE-targeted biopsy had nearly twice the detection rate of positive cores compared to systematic biopsy. When systematic biopsies were combined with targeted biopsies, it is seen that an increase in prostate cancer detection rate was occurred.

-Further studies are needed to assess prostate cancers in other zones, particularly in the transition zone. For accurate studying the number of cases must be increased as well as the cutoff values for sextants and patients need to be validated in larger multicentric clinical studies.

### 5. Different fluorescence image based cancer diagnosis

Fluorescence lifetime imaging (FLIm) offers a pathway for future in vivo applications to identify tumor infiltrations in the surgical bed which provides surgeons with additional information (reflecting the histopathology of the tissue) at the resection margin.

-Fluorescence confocal microscopy (FCM) provides a digital, microscopic, reliable diagnosis of prostate cancer with easy handling of specimens (without requiring staining and processing). These features makes FCM an affordable option across all surgical divisions, without the need for a dedicated setup. FCM also offers immediate sharing and reporting data by a remote pathologist via its real-time acquisition of digital images. If external validation confirms the findings, FCM technology could provide with high accuracy the discrimination between cancerous and normal tissue which could be applied to various settings and diseases, enhancing patient care during biopsies for suspected cancer.

### 6. Hyperspectral imaging and cancer diagnosis

-Utilizing the next generation of hyperspectral imaging systems as a real-time visualization tool during the process of the tumor resection will guide neurosurgeons via offering several TMD maps per second.

-The future intraoperative hyperspectral imaging system aims to utilize snapshot hyperspectral cameras capturing approximately ten images per second, allowing for real-time tracking of dynamic tissue changes. Steps such as generalizing results, optimizing algorithms, validating findings, expanding the image database, and enhancing the acquisition system are crucial for moving forward.

### 7. DEEP LEARNING BASED BREAST CANCER DIAGNOSIS WITH PET/MRI IMAGE

For different clinical applications and real-time diagnosis, CNNs can enhance the accuracy and they have been evaluated with using with PET/CT and MRI for assessing the pathological response of neoadjuvant chemotherapy in advanced breast cancer.

All in all, the conclusion emphasizes the need for innovations in hardware accelerators and AI algorithms to address challenges in synthetic and imbalanced datasets for accurate real-time cancer diagnosis.

REFERENCES.

Table I. challenges of Different Imaging methods for realtime cancer diagnosis and and Future works

| DIFFERENT IMAGING METHOD FOR CANCER DIAGNOSIS | CHALLENGES AND FUTURE WORKS |
|---|---|
| ULTRASOUND IMAGE FUSION | There are still challenges to address to use augmented reality clinically in surgeries involving soft-tissue organs. Since surgical targets and surrounding anatomies can move or deform during surgery due to respiration, heartbeat, or surgical manipulation. |
| | -Robot-assisted targeting provides enhanced accuracy. Incorporating 3D real-time digital registration and targeting is crucial for supporting focal therapy. |
| | -Studies assessed the accuracy and benefits of real-time image fusion in liver indicate that procedure accuracy may be affected by factors such as respiration, patient positioning, and the choice of landmarks for coregistration. |
| | -The fusion of US with CT or MRI in neurosurgery has demonstrated promising outcomes, despite the primary challenge posed by brain shifting due to dural opening. This can be a hurdle for achieving precise coregistration. |
| | -Although liver US is an essential tool in clinical medicine, it has limitations when it comes to diagnosing or performing interventional procedures on focal liver lesions. |
| DIFFERENT REAL-TIME OPTICAL IMAGING BASED CANCER DIAGNOSIS | - In a multicenter setting, real-time optical assessment cannot eliminate the possibility that features would have been scored differently by other endoscopists.<br><br>-The training session where definitions were discussed may have reduced significant interobserver variation. Additionally, variation in rater assessment could be beneficial for generalizing the results to real-world practice. However, the real-time setting led to some LNPCPs being biopsied before optical assessment. It is important to consider the evaluation of individual endoscopists' accuracy, as the number of LNPCPs assessed per endoscopist was too small to provide individual data. This limitation also prevented exploration of whether factors such as the type of scope used influenced accuracy.<br><br>-The number of T1 colorectal cancers was still limited, so the score chart was restricted to discriminating T1 colorectal cancers from non-invasive polyps without considering different submucosal invasion depths. Future studies should evaluate this by possibly updating the risk score with other features. The limited number of T1 colorectal cancers unavoidably led to less precise estimates.<br><br>-One limitation of the research conducted with PR-OCT is the ex vivo nature of all imaged specimens, as the human in vivo environment is likely more complex. The system was tested on a very limited number of other abnormalities, and the ability to differentiate adenomatous from hyperplastic polyps would have a significant clinical impact. Since most biopsy-proven hyperplastic polyps do not undergo surgical resection due to their benign nature, no patients with incidentally found hyperplastic polyps were encountered. It is necessary to test PR-OCT's ability to differentiate these two types of polyps in future in vivo patient studies. Additionally, the system was tested on a very limited number of tumors that had previously received radiation and chemotherapy treatment, resulting in a limited number of specimens. These treated lesions may require more categories in the PR-OCT classification design. It is also worth mentioning that a small training sample size can reduce the prediction power for abnormal lesions.<br><br>-Future efforts will include both hardware and software integration of PR-OCT into the endoscope, fine-tuning the network, and evaluation in the in vivo setting.<br><br>-The CADx tool cannot identify sessile serrated polyps, a recently recognized polyp type with likely neoplastic potential. Another limitation is the learning curve of the colonoscopists during the study period, as the prospective study design may contribute to underestimation of the CADx performance.<br><br>-Future cost-effectiveness studies in colonoscopy with CADx may investigate whether the extended procedure time is justified by the potential benefits of reduced polypectomies. |

,

| | |
|---|---|
| REAL-TIME IN VIVO CANCER DIAGNOSIS WITH DIFFERENT SPECTROSCOPY TECHNICS | -To advance the Raman technique for early cancer detection, collaborative efforts are needed among scientific teams comprising biophysicists, clinicians, and biomedical engineers. Further validation of Raman spectroscopy's clinical utility is essential. Additionally, the adoption of new technologies as they emerge is crucial for enabling real-time, in vivo diagnosis across various types of cancer.<br><br>-Removing noise and artifacts from real-time Raman spectra of live tissues remains a challenge, as they can interfere with interpretation in pathology. Novel designs of Raman fiber optic probes and data handling algorithms are necessary to enable fully automated spectral analysis.<br><br>-Future investigations should focus on addressing the challenge of fusing images with large anatomical variation due to organ deformation before the widespread use of fusion imaging in managing gastrointestinal malignancies.<br><br>-Conducting a multicenter trial would facilitate the recruitment of participants from a broader population. However, the method for correlating tumor location on the specimen was constrained by the histopathology protocols.<br><br>-H&E slides had to be used in conjunction with images of the sliced tissue and the entire specimen to manually label tumor areas. This approach may have resulted in mislabeling some data points on the border between the tumor and healthy tissue.<br><br>-Furthermore, samples labeled as tumor but later confirmed to be regressed tumor or fibrotic tissue were excluded. Since distinguishing between these types of tissue is challenging for surgeons based solely on tactile and visual assessments, future studies would greatly benefit from classifying these tissues.<br><br>-Integrating DRS technology into the surgical workflow for tumor mapping can be facilitated through augmented reality or an optional heatmap overlay added to a laparoscopic or robotic camera feed, which only minimally deviates from the standard plan of care. A concurrent in vivo validation of the real-time tracking technology and classification is planned, requiring adaptation of the probe tracking system. Advanced deep learning neural networks can assist in accurate probe tip location detection and tracking in DRS. Correlating real-time tissue classification in DRS with clinical outcomes such as recurrence, positive resection margin rate, or overall survival will be necessary to demonstrate the clinical utility of this technology before its potential clinical adoption. |
| ELASTOGRAPHY BASED CANCER DIAGNOSIS | -There is a risk that cancers may be missed or underestimated in terms of their size or Gleason score, which could explain the high false-positive rate of SWE. Using biopsy results as the reference standard also has limitations, including the lack of precise correlation between prostate cancer location and SWE estimates. This means that stiff areas detected by SWE may not always correspond to actual cancerous tissue, leading to potential misdiagnoses.<br>-Additionally, there is a need to address the lack of data on inter- and intra-observer variability, as this can impact the accuracy of prostate SWE imaging. To improve accuracy, operators should release pressure on the transducer when a stiff area is detected and image the lesion from different angles to avoid misdiagnosis.<br>-Moving forward, the focus should be on objective comparisons between systematic and RTE-targeted biopsy approaches to detect clinically significant prostate cancer. Employing deep learning systems for real-time analysis and diagnosis under complex conditions is crucial for clinical AI applications. Multimodal data processing has shown to enhance the robustness of AI-aided diagnosis. |
| DIFFERENT FLUORESCENCE IMAGE BASED CANCER DIAGNOSIS | -Further improvements are needed to enable the detection of microscopic lesions in the surgical field, which might otherwise be missed, and to possibly prevent damage to nearby uninvolved vital structures such as nerves. To ensure the success of FLE based on PH as a practical medical device, two key issues need to be addressed: in vivo pH measurement and intravenous administration for fluorescein. These issues should be discussed and resolved by comparing in vivo pH measurement between the proposed system and a commercial pH measurement with intraoperative intravenous fluorescence injection in further studies. |
| HYPERSPECTRAL IMAGING AND CANCER DIAGNOSIS | -The HR-MMWI system's functionality can be improved by integrating all imaging antennas and their corresponding circuits into a single framework using microwave integrated circuit technology.<br>-Utilizing hardware acceleration platforms such as GPUs or FPGAs to implement the full brain cancer detection algorithm is possible. This implementation should explore the design space to achieve the optimal balance between real-time execution, memory usage, and power dissipation using heterogeneous platforms. |
| PET/MRI IMAGE DEEP LEARNING BREAST CANCER | Further studies are necessary to enhance the model's ability to inform clinical treatment decisions. |